\documentclass[twocolumn]{aastex631}

\newcommand\lsim{\mathrel{\rlap{\lower4pt\hbox{\hskip1pt$\sim$}}
\raise1pt\hbox{$<$}}}
\newcommand\gsim{\mathrel{\rlap{\lower4pt\hbox{\hskip1pt$\sim$}}
\raise1pt\hbox{$>$}}}

\usepackage{color}

\shorttitle{Wide binaries in the galactic center}
\shortauthors{Michaely and Naoz}
\graphicspath{{./}{figures/}}
\begin{document}

\title{New Dynamical Channel: Wide Binaries in the Galactic Center as a Source of Binary Interactions}

%% you desire. In v6.31 you can include a footnote in the title.

\correspondingauthor{Erez Michaely}
\email{erezmichaely@gmail.com}

\author{Erez Michaely}
\affiliation{Department of Physics and Astronomy, University of California,
Log Angeles, CA 90095, USA}
\affiliation{Mani L. Bhaumik Institute for Theoretical Physics, University
of California, Log Angeles, CA 90095, USA}

\author{Smadar Naoz}
\affiliation{Department of Physics and Astronomy, University of California,
Log Angeles, CA 90095, USA}
\affiliation{Mani L. Bhaumik Institute for Theoretical Physics, University
of California, Log Angeles, CA 90095, USA}

\begin{abstract}
The inner $500\rm pc$ in the galactic center is dense with stars and a dynamically hot environment. Here, we focus on wide binaries as a source of tidally or collisional interactions. These binaries were previously ignored as sources of binary interaction because they are too wide to have a close pericenter passage, or they will quickly become unbound (ionized) due to gravitational interactions with passing neighbors. However, we show that wide binaries tend to interact more frequently with neighboring stars due to their larger cross-section for gravitational impulse interactions. These interactions mainly torque the wide system, causing it to change its eccentricity. As a result, the eccentricity might be excited to sufficiently high values, causing the binary to interact at the pericenter. As a proof of concept, we present four channels of such interactions: binary main-sequence (MS), white-dwarf (WD) - MS, black hole - MS, and lastly, WD-WD. During Galaxy's lifetime, we predict tens of thousands of MS-MS interacting binaries that may form G2-like objects later appear younger than their environment. X-ray signatures and, perhaps, supernovae may result from thousands of WD-MS and  WD-WD interacting binaries from this channel. Lastly, we predict a few hundred BH-MS interacting binaries at the inner $500$~pc. 
%To be filled

\end{abstract}

\keywords{stars: kinematics and dynamics
 --- keyword2  --- keyword3}

\section{Introduction} \label{sec:intro}

It is well known that almost every galaxy, our own Milky Way (MW) included, host a supermassive black hole (SMBH) at it's center, with masses between $10^6M_\odot - 10^9M_\odot$ \citep{Ghez1998,Ghez2000,2008Ghez,Gillessen2009,Genzel2003,Genzel2010,Ferrarese2005,Kormendy2013}. The SMBHs are surrounded by dense regions of stars and compact objects \citep[e.g.][and others]{Ghez2003,Ghez2005,Alexander2005,Hopman2009,Chu2018ApJ}. In the context of the stellar dynamics in the center of a galaxy, we consider here two distinct regions. One is the inner region which is completely dominated by the gravitational potential of the SMBH, and the other is an extended region all the way toward the Galactic field.

The inner region, called "the sphere of influence," extends to a radius where each object is dominated by the SMBH gravitational potential.   
%. In this region every star, binary or compact object is reacting primarily to the SMBH. 
In most galaxies within the sphere of influence reside a very dense star cluster named nuclear star cluster  \citep[][]{Ghez2003,Ghez2005,Alexander2005,Hopman2009,Chu2018ApJ,Neumayer2020}, though it can extend to larger radii. These NSCs are the densest star clusters in the Universe, hence host to very interesting dynamical scenarios \citep[e.g.,][and others]{Magorrian1999,Merritt2004,Hopman2005,Antonini2010,Antonini2011,Prodan2015,Aharon2016,Sari2019,Linial2022,Naoz2022,Rom2023}{}{}. The mass within the sphere of influence is approximately two times the mass of the central SMBH. The size of the sphere of influence is given by $r_h\approx GM_{\rm SMBH} /\sigma_0^2$, where $M_{\rm SMBH}$ is the SMBH mass and $\sigma_0$ is the velocity dispersion. In the MW $\sigma_0\approx75\rm{kms^{-1}}$ \citep{Gebhardt1996,Gebhardt2000,Gebhardt2000b} this corresponds to an estimated radius of $\sim 3.1\rm{pc}$ enclosing millions of stars.

Beyond the radius of influence, the density of stars remains high and slowly decreases as it converges with the galactic bulge \citep{Ballero2007,Athanassoula2017,Barbuy2018} and further away from the Galactic field.
%is beyond this radius of influence, where
At this regime, the gravitational potential is dominated by both the SMBH and the stellar mass enclosed. % a dense region of stars, albeit less dense than the NSC, extended all the way to the field of the galaxy. 
In the MW this corresponds to a maximal radius of ~$500-1000\rm{pc}$ with ~$10^9-10^{10}$ of stars.  Aside from stars and compact objects inside the galactic bulge, there is a presence of unusually high concentration of molecular gas. These molecular gas clouds are conducive to star formation and the birth of new stars in the bulge and this central molecular zone \citep[CMZ][e.g.]{Mills2017,Langer2017,Sormani2020}. 

In the context of this work, we largely neglect the gas contribution to the dynamics because the gas density is still low and will have a negligible effect on the dynamics. On the other hand, the star formation that takes place in the CMZ may give rise to young stars and binary stars. Particularly, in the field, binaries and higher multiplicity systems are very common \citep{Tokovinin1997,raghavan2010,Sana2014,Moe2016}. Further, binaries and high multiples white dwarfs have been shown to be prevalent in the local neighborhood \citep[e.g.,][]{Kraus2009,El-Badry2021,Heintz2022,Shariat2023}.

Additionally, the detection of a few binaries in the MW's NSC \citep[e.g.,][]{Ott1999,Martins2006,Pfuhl2014} indicated a broader binary population in the NSC. Furthermore, other observational studies point out that the OB binary fraction in the NSC is comparable to the fraction found in young massive stellar clusters \citep[e.g.,][]{Ott1999,Rafelski2007}, and the fraction of Eclipsing young OB binaries in the galactic center is consistent with the local fraction observed \citep{Chu2018ApJ,Gautam2019}. The theoretical
study \citep{Stephan2016} showed that the binary fraction is $\sim 70\%$ for the population from the most recent star formation episode, 6 Myr ago, in the NSC  \citep[e.g.,][]{Lu2013}. Lastly, the peculiar configuration of the young stellar disk in the galactic center suggests it hosts a high fraction of binaries \citep{Naoz2018}. 

%The presence of binaries is crucial to the formation of many astrophysical phenomena, such as supernovae, X-ray sources, GW sources, and others {\sn{CITE}}. Hence, in this {\it letter}, we account for the presence of binaries in the galactic bulge and study the interaction between the binaries and other stars in the region extending all the way to the galactic bulge.

In recent years, it was well established that wide systems ($>1000\rm AU$), of either binaries or triples, in the field of their host galaxy are a rich source of binary interaction \citep{Kaib2014,Michaely2016,michaely2019,michaely2020,Michaely2020b,Michaely2021,michaely2021a,Raveh2022,michaely2022,Grishin2022,Rozner2023}. Notably, while the galactic field is generally considered collisionless for many dynamical processes, it is not the case for wide binaries. Wide binaries, due to their large geometrical cross-section, which is proportional to their semi-major axis (SMA), interact frequently with flyby field stars. These interactions are well modeled by the impulse approximation, and their rate is given by the standard $\Gamma=n_* \sigma_{\rm{geo}} v_{\rm{env}}$. Where $n_*$ is the local stellar density, $\sigma_{\rm{geo}}$ is the geometric cross-section, and $v_{\rm{env}}$ is the relative velocity between the binary and the flyby of the encounter, which we take to be the local velocity dispersion.

In this {\it letter} we study the interaction between wide binaries in the galactic center and flyby stars. Notably, these interactions can naturally result in observational signatures. Particularly here, we explore four different dynamical channels as proof of the importance of the proposed dynamical process. These channels involve main-sequence stars, white dwarfs, and black holes, see below. 

%The first channel we explore is the binary main-sequence (BMS), wide BMS could be driven to interact significantly at pericenter and produce several interesting outcomes: blue-straggler, stars that appear bluer than other stars with the same luminosity in their environment \citep[][]{Leonard1989}, a common envelope evolution, a binary evolutionary stage where the two stars cross each other Roche limit and share an envelope \citep{Webbink1984}, amorphous cloud formation and a direct collision \citep{Kaib2014} . 

%The second channel is the interaction between a black-hole (BH) and a main-sequence (MS) star. After the formation of the BH, a wide BH-MS binary could be kicked to sufficiently high eccentricity in order to dissipate energy and angular momentum to create a compact binary which upon the evolution of the MS to a giant might produce an X-ray source \citep{Michaely2016}.

%The third channel we explore here is the interaction between a white-dwarf (WD) and a MS. These very common binaries upon interaction might produce:  cataclysmic variable \citep{Knigge2011}, novae eruptions \citep{Chomiuk2021}, Type Ia supernova progenitor \citep{Maoz2014,Livio2018,Soker2019}

%The fourth and last channel we study it possible outcomes is the double WD binary (DWD). When two WDs could be kicked into very high eccentricities these are the possible interesting outcomes: Type Ia supernova progenitor (if the double degenerate scenario works or the direct collision scenario \citep{Iben1984,Webbink1984}), GW LISA source \citep{Korol2018}.

This paper is organized as follows: In section \ref{sec:Physical Picture} we present the dynamical scenario qualitatively and quantitatively. In section \ref{sec:Binaries_in_the_inner} we describe the population model used in our calculations, in section\ref{sec:Results}  we present the four channels the rates, and the distribution of the systems of interest. In section \ref{sec:Discussion} we discuss the results, and we summarize in section \ref{sec:Summary}.

\section{\label{sec:Physical Picture}Physical Picture}

%\subsection{The Collisional Nature of Wide Binaries: qualitative description\label{subsec:qualitative}}

As mentioned in Sec. \ref{sec:intro} the galactic field \textit{is} collisional for wide systems \citep{Kaib2014,Michaely2016,michaely2019,michaely2020,Michaely2020b,Michaely2021,michaely2021a,Raveh2022,michaely2022}. Expanding on this body of work we describe the interaction of wide binaries in the inner $500 \rm{pc}$ of the galactic center. We note that in this manuscript we focus on binaries, we leave the treatment for triples to future study.

In what follows we describe, in a qualitative manner, the physical picture of the dynamical scenario, namely the gravitational interaction between wide systems and random flyby stars in the galactic center. The galactic center is a collisional environment, in which the soft-hard binary \citep{Heggie1975} is determined by the average kinetic energy of the stars. We focus on the soft binaries, hereafter wide binaries. Wide binaries interact randomly with flyby stars, the encounter velocity is determined by the local velocity dispersion. The gravitational interaction is impulsive by nature, e.g. the interaction timescale $t_{{\rm int}}\equiv b/v_{{\rm enc}}$ is much shorter than the orbital period of the soft binary, $P$. Where $b$ is the closest approach of the flyby to the binary's center of mass. It was shown in the literature that these interactions change the binary orbit characteristics, namely the orbital energy (decrease/increase the SMA), but more significantly, torque the system and change the binary eccentricity $e$ \citep[e.g.,][]{Lightman1977,Merritt2013}. 

Such consecutive interactions change the orbital eccentricity and at times might excite the eccentricity to critical values that lead to interaction at pericenter. This process occurs at the same time with the "ionization" process of soft binaries. Due to random interactions of these soft binaries with flyby stars they gradually are "ionized", decreasing the number of available binaries to interact at pericenter due to the process described here. The timescale for the ionization is given by the average halt-life time of the binary is  \citep{Bahcall1985}:

\begin{equation}
t_{1/2}=0.00233\frac{v_{{\rm enc}}}{Gm_{p}n_{*}a}.\label{eq:t_half_life}
\end{equation}
As a result of the random interaction, we expect to find an increasing pericenter interaction probability as a function of time, due to the random nature of the encounters. However, the ionization of the binaries leaves the total binaries number a decreasing function of time, as the total number of binaries available for interaction decreases. We note here that as the ionization probability is a function of the SMA, we expect the contribution of the widest systems only happens at early times since formation. 

%In the next subsection we present a qualitative treatment to this scenario, accounting all the physical ingredients describe thus far.

%\subsection{The Collisional Nature of Wide Binaries: quantitative description\label{sub:quantitative}}

We Consider an ensemble of isotropic wide binaries in the galactic center, each is comprised of two objects with masses $m_1$ and $m_2$, and the total (reduced) mass is denoted by $M \left(\mu\right)$. All binaries have the same SMA $a$, and a thermal distribution of orbital eccentricities, $f(e)de=2ede$. We define a length scale called the interaction distance to be $d$, this length scale is set by the type of interactions we are interested in the pericenter.

As an illustrative example in \citep{Michaely2016}, we were interested in calculating the formation of an X-ray source from an initially wide BH-MS binary, hence we set $d$ to be the tidal radius for which the binary will lose enough orbital energy and eventually become an X-ray source, specifically we set $d=5R_*$.

Once one sets the interaction distance $d$, the critical eccentricity, $e_c$, for which each pericenter distance, $q=a(\left(1-e\right)$ is equal or smaller than $d$ is given by

\begin{equation}
e_{{c}}=1-\frac{d}{a}.\label{eq:e_crit}
\end{equation}

Next, we calculate the loss cone fraction, $F_q$, namely, the fraction of systems, given thermal distribution of eccentricities, that satisfies $q\le d$, which is equivalent to $e\ge e_c$ 

\begin{equation}
F_{q}=\int_{e_{c}}^{1}2ede=1-e_{c}^{2}=\frac{2d}{a}.  \label{eq:F_q}
\end{equation}
In the last Equation, we used the approximation $d/a\ll1$.

Now, we are ready to introduce the interaction with the environment, specifically with flyby stars. The environment in which the ensemble of wide systems resided is characterized by its local stellar density, which is a function of distance from the center of the host galaxy, $n_*\left(r\right)$, the local velocity dispersion also as a function of distance, and the average mass of stars. We model the stellar density function by the following  \citep{Genzel2010,Rose2020}
\begin{equation}
n_*\left(r\right)= \frac{M_{\bullet}}{2\pi r_h^3 m_{\rm mean}} \left(\frac{r}{r_h}\right)^{-\alpha} \ ,\label{n_center}
\end{equation}
where $r_h=GM_{\bullet}/\sigma^2_0$ is the radius of influence, and $\sigma_0$ is the velocity dispersion in the radius of influence, set to be $\sigma_0=75\rm{kms^{-1}}$. The SMBH mass is $M_{\bullet}$ and we set it to be $M_{\bullet}=4\times10^4M_{\odot}$.
The velocity dispersion is a decreasing function of $r$ from the radius of influence until the boundary of the field; we adopt the following toy model
 \begin{equation}
\sigma \left(r\right)= \rm{Max} \left(\sigma_0 \left(\frac{r_h}{r}\right)^{1/2} ,\sigma_{\rm field}\right) \ .
\label{eq:sigma_center}
\end{equation}
The flyby mass is taken to be $m_p=1M_\odot$, $\sigma_{\rm field}=25\rm{kms^{-1}}$ and we use $\alpha=2$.

Each interaction with the perturber changes the relative velocity between the binaries components (e.g., see eq.(7)) in \cite{Michaely2016}. One can calculate the smear cone, $F_s$, which is the fraction of phase space to which the initial relative velocity could be kicked into after a single passage of the perturber as a function of $b$

\begin{equation}
F_{s}=\frac{\pi\theta^{2}}{4\pi}=\frac{27}{4}\left(\frac{m_{p}}{M}\right)^{2}\left(\frac{G\mu}{av_{{\rm enc}}^{2}}\right)\left(\frac{a}{b}\right)^{4} \ .\label{eq:F_s}
\end{equation}

For the loss cone to be continuously filled, two conditions must be fulfilled. First, the interaction between the perturber and the wide binary should be sufficiently strong in order to fill the loss cone. This condition is achieved when $F_s\ge F_q$. The second, the replenishment rate into the loss cone, is equal to the rate of losing binaries from the loss cone, specifically the binary dynamical timescale. The rate of replenishment is given by the rate of interaction, which is
\begin{equation}
f\equiv\frac{1}{t_{{\rm enc}}}=n_{*}\sigma\left(b\right) v_{{\rm enc}} \ . \label{f_t_enc}
\end{equation}
The condition $F_s\ge F_q$ is actually a condition of the closest approach $b$, 
\begin{equation}
b_{\rm kick}^2 \le v_{\rm enc}^{-1}\left(\frac{27}{8}\frac{m_p^2}{\mu}\frac{G\mu\pi}{d} a^{4} \right) ^{1/2} \ . 
\label{eq:b_kick}
\end{equation}

\begin{figure*}
\includegraphics[width=1.08\columnwidth]{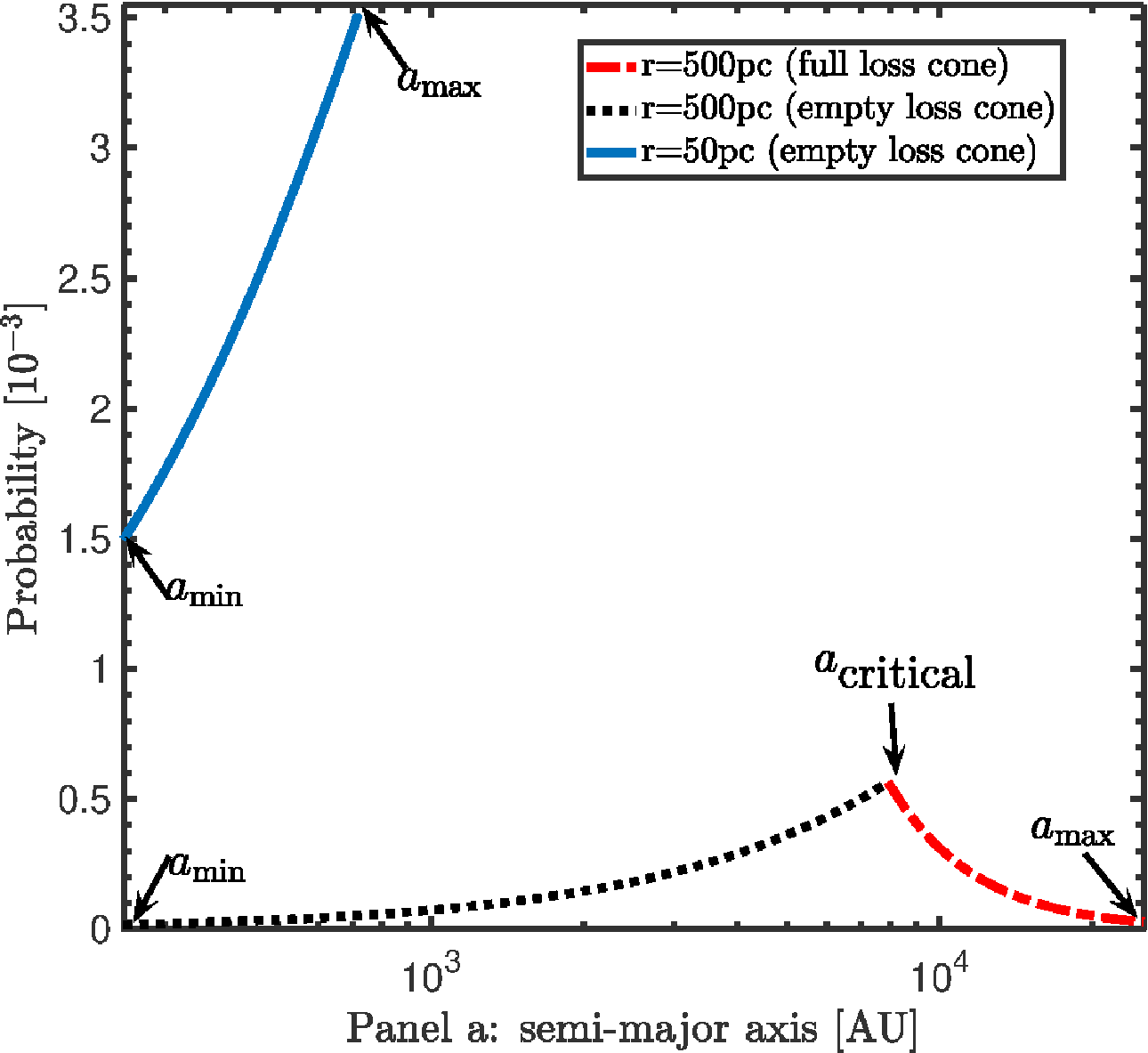} \quad \includegraphics[width=1\columnwidth]{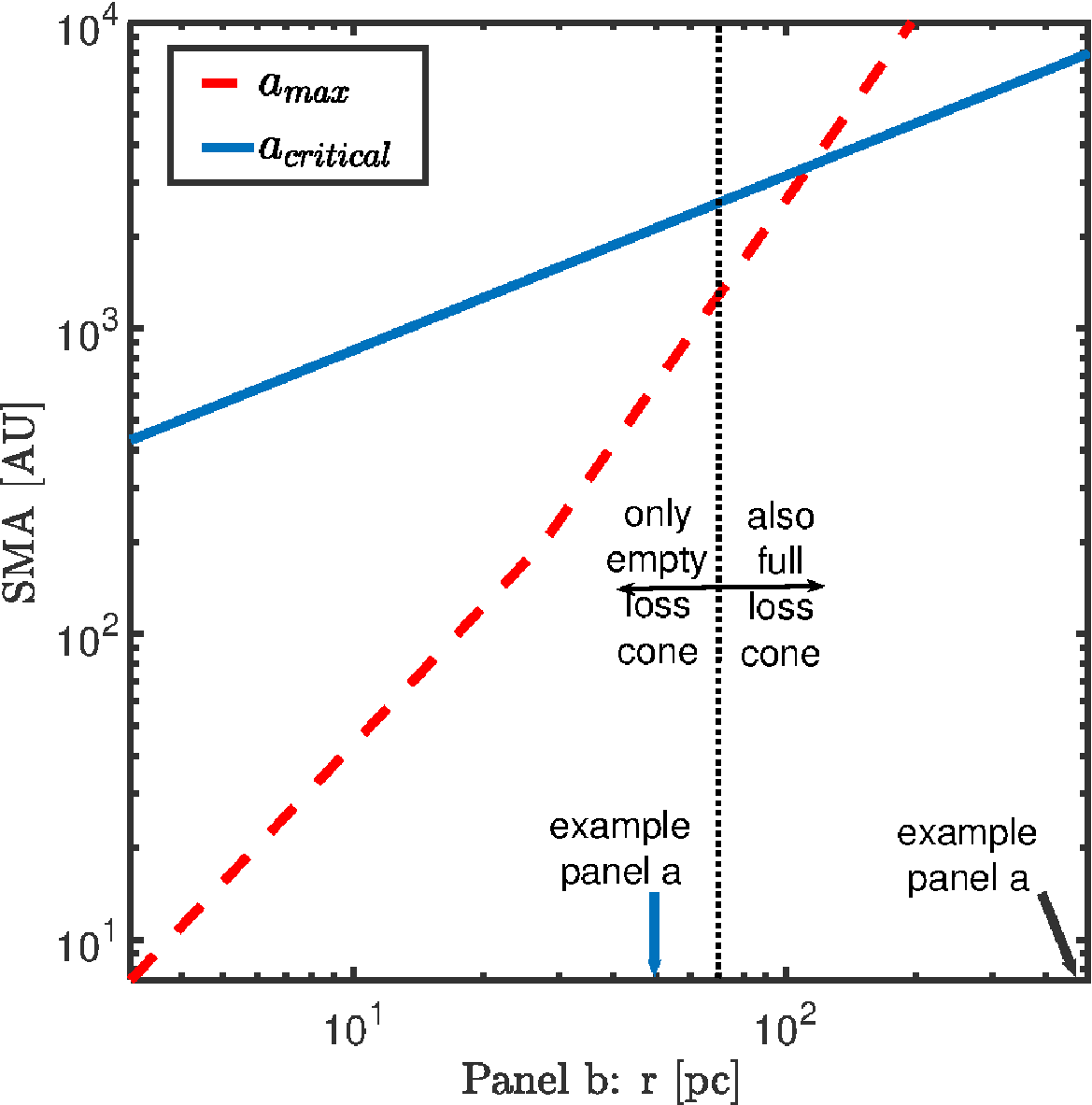}

\caption{\label{fig:The-merger-probability}
Left panel (panel a): The probability for BMS interaction as a function of the binary SMA. The probability is calculated at $t=50\times10^6 \rm{yr}$ for two MS stars each with mass of $1M_\odot$. For two locations in the galactic center. The blue solid line is at $r=50\rm{pc}$ and the black dotted line and the red dashed line are at $r=500\rm{pc}$.
Right panel (panel b), the BMS case. The blue solid line is the critical SMA from eq. (\ref{eq:a_crit}). The red dashed line is $a_{\rm max}$ the upper cut-off of the BMS binaries see eq. (\ref{eq:a_max}). For binaries inside $\lesssim 200\rm pc$ the $a_{\rm crit} > a_{\rm max}$. Namely, only the empty loss cone contributes to the rates, and for $r\gtrsim 200\rm pc$ an added contribution from the full loss cone.}
\end{figure*}

Now, we are set to find the critical SMA from which the loss cone is continuously full, under the impulse approximation, as a function of the local environment characteristics. Thus, for $F_{s}=F_{q}$ and $1/t_{{\rm enc}}=n_{*}\sigma v_{{\rm enc}}$, we find: 
\begin{equation}
    a_{{\rm crit}}=\left(\frac{2}{27\pi^4}\frac{\mu M}{m_p^2}\frac{d}{n^{2}_*}\right)^{1/7}\label{eq:a_crit} \ .
\end{equation}
Binaries with SMA equal to the $a_{{\rm crit}}$ have the highest probability of interacting at the pericenter. Binaries with SMA smaller than $a_{{\rm crit}}$ are harder to perturb. Therefore, the probability rate decreases. Binaries with SMA larger than $a_{{\rm crit}}$ also decrease in probability due to the smaller loss cone, additionally, these binaries are more likely to become unbound (ionized) during their life-time. This interplay leads to a sharp transition in the merger probability, as highlighted in panel a of Figure \ref{fig:The-merger-probability}.

In other words, the full loss cone regime is for all systems with $a>a_{{\rm crit}}$. The loss rate of systems from the full loss-cone is governed by the timescale where the binaries are lost from the cone. This timescale is the orbital period, $P$. Hence, ($\dot{L}_{{\rm full}}$), is given by 
\begin{equation}
\dot{L}_{{\rm full}}=\frac{F_{q}}{P}. \label{eq:Loss_rate-1}
\end{equation}
On the other hand, tighter binaries are less susceptible to change due to a fly-by, this is evident from equation (\ref{eq:F_s}). Therefore for $a<a_{{\rm crit}}$ we expect that the loss cone will not be full all the time, in this ``empty loss cone'' regime the loss rate depends on the rate of orbits being kicked into the loss cone is just the interaction rate, $f$:
\begin{equation}
\dot{L}_{{\rm empty}}=F_{q}f=\frac{F_{q}}{t_{{\rm enc}}}
\end{equation}
From the critical SMA we can calculate the merger probability for each regime, for $a<a_{{\rm crit}}$ (empty loss cone) and for $a>a_{{\rm crit}}$
(full loss cone). We note here that by considering the critical SMA we implicitly include the dependency on the impact parameter, $b$.

Next, we calculate the loss probability which is time and space-dependent for the two regimes, the empty and full cones.

The loss cone $F_q$ is the fraction of systems that are lost from the ensemble, hence $\left(1-F_{q}\right)$ represents the fraction of binaries
that survive after a single flyby with the relevant timescale. Therefore, $\left(1-F_{q}\right)$ is a monotonically decreasing function of time. Hence, the surviving fraction of binaries after time $t$ and the relevant timescale for the empty loss cone $f$ is $\left(1-F_{q}\right)^{tf}$ \citep[e.g.,][]{Kaib2014,Michaely2016,michaely2019}. Therefore the probability for a wide binary interaction, which complements this expression to unity is
\begin{equation}
L_{a<a_{{\rm crit}}}=1-\left(1-F_{q}\right)^{tf}\label{eq:empty_Prob}
\end{equation}
where $t$ is the time since the birth of the binary. As one can expect
the probability only depends on the size of the loss cone and the
rate of interactions. For the limit of $tfF_{q}\ll1$
we can expand Equation (\ref{eq:empty_Prob}) and take the leading
term, to find the loss probability to be approximated by 
\begin{equation}
L_{a<a_{{\rm crit}}}=tf(b)F_{q}.\label{eq:empty_prop_approx}
\end{equation}

For the full loss cone regime, the limiting factor is not the value of $f$, but rather the orbital period $P$. Therefore, the full expression for the loss probability for $a>a_{crit}$ is
\begin{equation}
L_{a>a_{{\rm crit}}}=1-\left(1-F_{q}\right)^{t/P}.\label{eq:full_prob}
\end{equation}
 For the limit $F_{q}\cdot t/P\ll1$ we can approximate the probability by 
\begin{equation}
L_{a>a_{{\rm crit}}}=tF_{q}~\frac{1}{P(a)}.\label{eq:full_prob_apx_express}
\end{equation}

The probabilities calculated up to here ignore the ionization of the binaries. To account for this process, we use the half-life time approximate relation given by \citep{Bahcall1985}, see eq. (\ref{eq:t_half_life})

Correcting for this we get for Equation (\ref{eq:empty_prop_approx})
and Equation (\ref{eq:full_prob_apx_express}) the following approximations:
\begin{equation}
L_{a<a_{{\rm crit}}}=\tau f(b)F_q\left(1-e^{-t/\tau}\right)\label{eq:ionization empty}
\end{equation}
and 
\begin{equation}
L_{a>a_{{\rm crit}}}=\tau \frac{F_q}{P(a)}\left(1-e^{-t/\tau}\right),\label{eq:ionization full}
\end{equation}
where $\tau=t_{1/2}/\ln2$ is the mean lifetime of the binary. For a more detailed derivation see equations (11) and (12) in \citet{michaely2019}.

The formation rate in the galactic center at time $t$ is given by the following integral:
\begin{equation}
\Gamma=\int_{r_h} ^{500\rm pc}\int_{a_{\rm min}} ^{a_{\rm max}} L f_{\rm a} \left(a\right) dN \left(r\right) da,\label{eq:Gamma}
\end{equation}
where the first integral is on the position in the galactic center, and the second is on the SMA of the binaries. $L$ is given by equations (\ref{eq:ionization empty}) and (\ref{eq:ionization full}). $f_{\rm a}\left(a\right)$ is the fraction of wide systems from the underlining distribution of the SMA. The total number of systems within the differential part of the galactic center $dr$ is $dN \left(r\right)$ given be eq. (\ref{n_center}).

In order to calculate this integral we need to define the physical characteristics of the population we consider. We do that in the next section.

%\subsection{Physical characteristics of the population\label{sub:phsical characteristics}}
\section{Binaries in the inner 500 \rm{pc}\label{sec:Binaries_in_the_inner}}
%In the previous subsection we modeled the interaction mathematically and derived the probabilities for interaction at pericenter for wide binaries as a function of their local environment.
%In this subsection we use the equations derived in the previous subsection and calculate the characteristics of the wide population of binaries.

Here, we apply the general equations presented above and apply them for the particular characteristics of the wide population of binaries at the inner $500$~pc. 

%\textbf{Upper SMA cut-off}. 
\subsection{Upper SMA cut-off}
As mentioned before the galactic center is dense to the point the half-life time of binaries is shorter than Hubble time. As a result, the expected SMA distribution for binaries changes as a function of time. In the galactic center, we calculate the upper cut-off of the SMA from the evolution time and the ionization time. For example, if we are interested in white dwarf binary (see section \ref{Sec:WD-WD}), we set the evolution time, $t_{\rm evolv}$ to be the time it take to form the white dwarf. This timescale corresponds to a SMA with a similar ionization timescale. This is the upper cutoff of the population we consider. Specifically, 
\begin{equation}
a_{\rm max} \left(t_{1/2}\right)= \frac{0.00233v_{\rm enc}}{Gm_pn_*t_{1/2}}  \ . \label{eq:a_max}
\end{equation}

%\textbf {Lower SMA cut-off}.
\subsection{Lower SMA cut-off}
%In order to be conservative 
We restrict our treatment to the impulsive approximation. Hence, we introduce a lower cutoff to the SMA. The typical kinetic energy per unit mass is higher at the galactic center; therefore, systems are regarded wide for much more compact binaries. We set the minimal value of the SMA to be such that the Keplerian velocity of the binary will be equal to the local velocity dispersion times a numerical factor, $\beta=5$. This gives the lower cut-off of the SMA distribution, namely:
\begin{equation}
a_{\rm min} = \frac{G\mu}{\beta\sigma}^2 \ .
\end{equation}

%\textbf {Wide binary fraction}. 
\subsection{Wide binary fraction}
In what follows, we calculate the fraction of binaries that reside in the span between $a_{\rm min}$ and $a_{\rm max}$. In order to compute this fraction at the beginning of the integration time, we assume the log-uniform distribution of the SMA, $a$ from the minimal value of $a_{\rm abs min}=2R_\odot$ that represents the most compact binary in the ensemble to $a_{\rm max}$ from eq. (\ref{eq:a_max}). The wide binary fraction is defined as the fraction of binaries the reside between between $a_{\rm min}$ and $a_{\rm max}$, namely
\begin{equation}
f_{\rm wide} = \int_{a_{\rm min}} ^{a_{\rm max}} c\times \frac{1}{a}da \ ,
\end{equation}
where $c$ is the normalization constant to the log-uniform distribution of the SMA. 
%Hence the 
%\textbf{Initial time}.

The last ingredient of the model is the formation time of the binary. The values depend on the specific binary we are studying. For example, for a binary MS, the formation time is the pre-MS lifetime, and for the binary WD, the formation time is the MS lifetime.

%Equipped with this modeling of the population we are set to use this model in order to calculate the interaction probability and unique signatures.

\section{\label{sec:Results}Binary channels and predictions}
This section presents four channels for interaction in the galactic center. For each channel, we describe the underlying population and present the outstanding results.

\subsection{\label{Sec:MS-MS}MS-MS and G2-like objects}

As the MS binary interacts at the pericenter, it loses energy and angular momentum and forms a compact binary. Several outcomes are plausible from this scenario; as a proof of concept in what follows, we only present them generally and reserve a more comprehensive study for future work. As mentioned above if the pericenter distance is sufficiently small the resulting outcome is may be either a merger or even a collision between the two MS stars. A merger BMS can lead to a merged object more massive and luminous than was is expected from its environment. These objects are known as blue-stragglers \citep{Leonard1989,shara1999,carney2001,Genzel2003,carney2005,perets2009,Kaib2014}. 

Another possible outcome from a merger of two MS stars is a transient event called a luminous red nova which is characterized by a prominent red color and a slow decaying light curve.\citep[e.g.][]{Kulkarni2007,Soker2012,Pejcha2016,Pejcha2016b,Pastorello2019}. 

Another possibility for an outcome is a common envelope evolution, a binary evolutionary stage where the two stars cross each other Roche limit and share an envelope \citep[e.g.][]{Webbink1984,ivanova2013,michaely2019a,Igoshev2020,Ropke2023,DiStefano2023}

We note, that this scenario is consistent with the idea that the G2-like objects host a stellar component \citep{Gillessen2012,Witzel2014,Prodan2015,Stephan2016,Witzel2017,Stephan2019,Gillessen2019,Ciurlo2020}. Collisions between two main sequence single stars have been proposed to be consistent with the G2 population closest to the SMBH \citep{Rose2023}. Here, we predict a population of G2-like objects at much larger distances from the SMBH. 

For this channel, we set the following parameters. The masses $m_1=1M_\odot$ and $m_2=1M_\odot$, represent the mass of a MS star. The interaction radius, $d=5R_{\odot}$, represents the distance in which tidal interaction is sufficient to dissipate orbital energy. From Kroupa IMF \citep{Kroupa2001}, we calculate the fraction of the primary to be an MS during the integration time \citep{Moe2016}. In order to do so, we limited the mass range to be between $0.5M_{\odot}-3M_{\odot}$.  The fraction of the primary to be in that mass span is $f_{m_1}=0.25$. Assuming a uniform distribution of the mass ratio $q\in \left(0.3,1\right)$, the fraction of the companion to be in the same mass range is $f_{m_2}=0.54$.  The binary fraction for this mass scale is $f_{\rm binary}=0.5$. The product of these fractions gives the fraction of MS-MS systems from a given population, $f_a=f_{m_1}\times f_{m_2} \times f_{\rm binary} \times f_{\rm wide}$ from the  We emphasis here for simplicity we represent all MS stars with a single mass of $1M_\odot$.

%In the context of this study, we count all of these possibilities as systems of interest. 

\begin{figure}
\includegraphics[width=1\columnwidth]{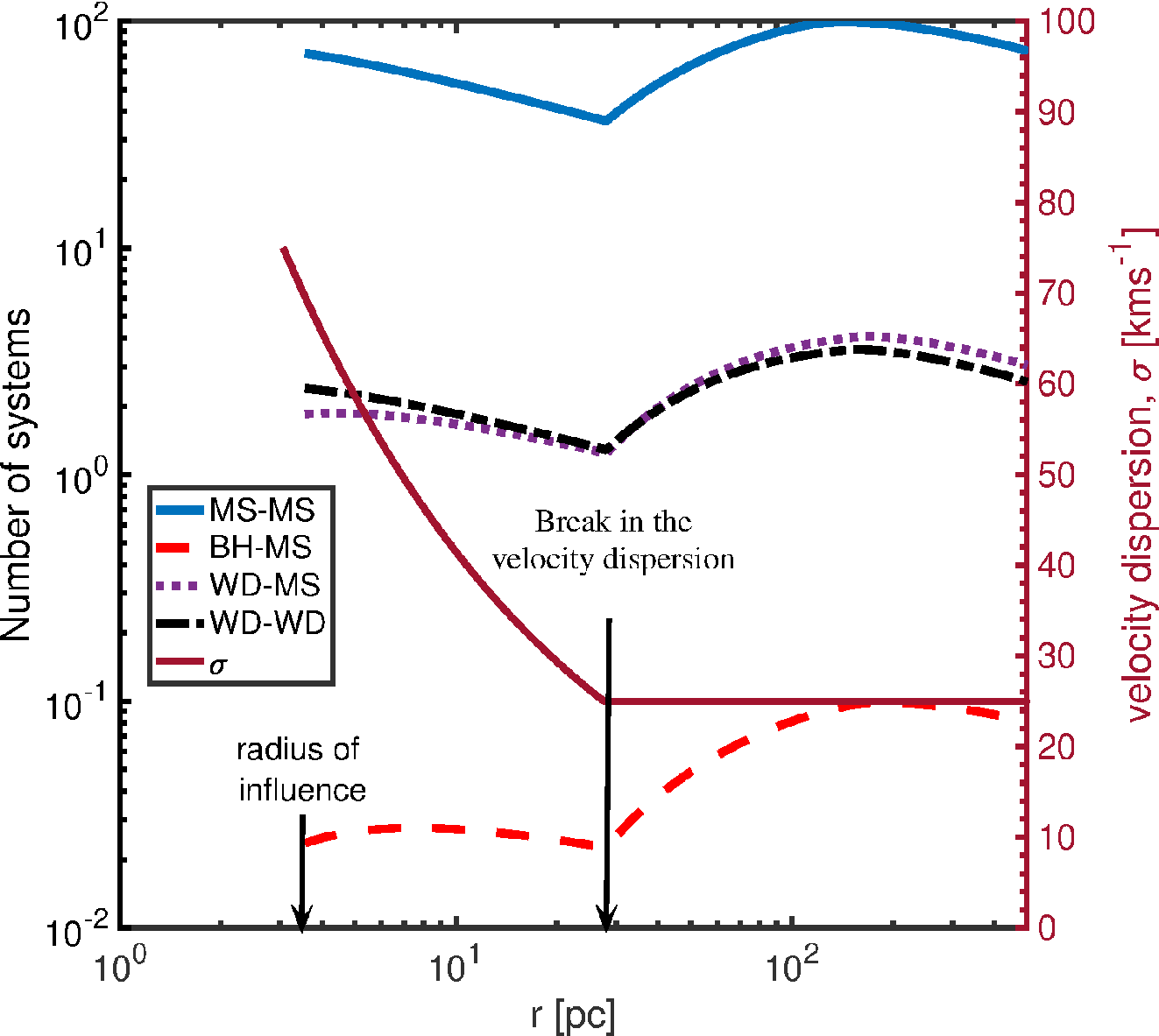}
\caption{\label{fig:All_N_LogR}Left Y axis: Total number of systems of interest as a function of distance from the SMBH for all 4 channels after Hubble time,$10^{10}\rm yr$ for the simplistic case of a single starburst at $t=0$. Blue solid line, denotes the BMS channel. The Red dashed line denotes the BH-MS channel. The purple dotted line represents the WD-MS channel. Black dashed-dot line is the DWD channel. All channels show similar structure, albeit with different total systems due to the different initial systems for each channel. The break at $r\approx 30\rm pc$ is due to the change in the velocity dispersion. The right Y axis shows the velocity dispersion, toy model (see Eq.~(\ref{eq:sigma_center}), as a function of distance from the SMBH, from the radius of influence, $r_h$ up to $500 \rm pc$. The peak all channel exhibit around $r\approx 200\rm pc$ is explained in section \ref{sec:Discussion} and in Figure \ref{fig:DTD_BMS_different_times}
}
\end{figure}

\begin{figure}
\includegraphics[width=1\columnwidth]{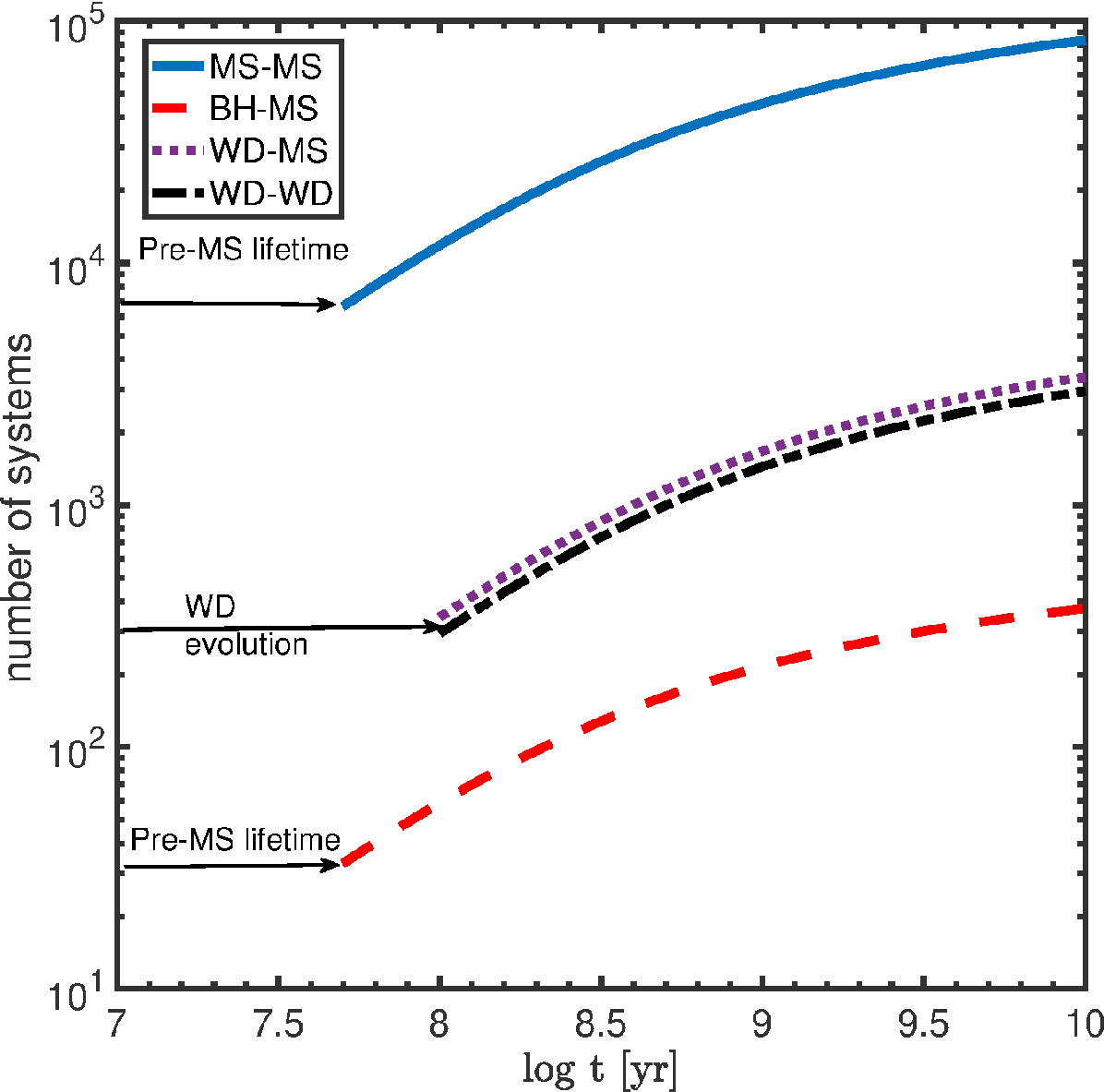}
\caption{\label{fig:All_N_logT}The total number of systems of interest for all channels. The color code is as Figure \ref{fig:All_N_LogR}. The total number of system is integrated over the galactic center. Specifically from the radius of influence, $r_h$ to the edge of the field, $r=500\rm pc$. Each channel has a different starting point due to the different time it takes the initial binary to evolve.}
\end{figure}

In Figure \ref{fig:All_N_LogR}, we present the number interacting MS-MS (BMS) %of %systems of interest from binary MS (BMS)
%channel
as a function of separation from the SMBH after Hubble time, $t=10^{10}\rm yr$, blue solid line. In Figure \ref{fig:All_N_logT}, we present the total number of systems of interest from BMS (in blue solid line) as a function of time (see labels in both Figures).

\subsection{\label{Sec:BH-MS}BH-MS and stellar transients}

The outcome of BH-MS interaction at the pericenter could lead to an X-ray source,  \citep[e.g.,][]{Michaely2016,Naoz2016,Hoang2022,Rose2023}.  Furthermore, in some cases these systems may lead to tidally disrupting the star \citep[e.g.,][]{Perets2016,fragione2019,Kremer2022,Kremer2023,Xin2023}.  For a high eccentric encounter,  these tidal disruption events may leave short optical or x-ray transient.  

In what follows we adopt $m_1=10M_\odot$ for the BH mass and $m_2=1M_\odot$ for the main sequence star.  Using Kroupa IMF  we estimate the fraction of these BHs from the population up to $500$~pc.
%For this channel, we adopt the following parameters. The masses $m_1=10M_\odot$ and $m_2=1M_\odot$ represent the mass of the primary BH and the secondary MS star, respectively. 
%The interaction radius, $d=5R_{\odot}$, represents the distance in which tidal interaction is sufficient to dissipate orbital energy.
%From Kroupa IMF we find the fraction of the primary to be a BH. 
We limited the mass range to be bigger than $20M_{\odot}$ at the zero-age MS.  Thus we find that the fraction of the primary is $f_{m_1}=2\cdot10^{-3}$. Assuming power law distribution of the mass ratio $q\propto m^{-2}$ from $\left(0.1,1\right)$ \citep{Moe2016}, the fraction of the companion to be an MS star is  $f_{m_2}=0.1$.  The binary fraction for this mass scale is $f_{\rm binary}=1$. We emphasize here for simplicity; we represent all BH and MS stars with a single mass of $10M_\odot$ for the BH and $1M_\odot$ for the MS star. Finally, the interaction radius, $d=5R_{\odot}$, represents the distance in which tidal interaction is sufficient to dissipate orbital energy. In Figure \ref{fig:All_N_LogR}, we present the number of systems of interest from the BH-MS channel as a function of the  distance from the SMBH after Hubble time, $t=10^{10}\rm yr$ in red dashed line. In Figure \ref{fig:All_N_logT} we present the total number of systems of interest in the red dashed line as a function of time.

\subsection{\label{Sec:WD-MS}WD-MS and CV/ novea}

WD-MD interactions can lead to a myriad of observational signatures, such as cataclysmic variable \citep[e.g.,][]{Muno2006,Muno2009,Knigge2011}, novae and other transients \citep[e.g.,][]{Shara1977,Shara1978,Shara1986,Webbink1987,Chomiuk2021}, Type Ia supernova progenitor (for the single degenerate scenario, e.g., \citet{Raskin2009,Raskin2010,Rosswog2009,Hawley2012,soker2019}, but see \citet{Loren-Aguilar2010}).  for the purposes of this work, we are interested in the general electromagnetic signature and leave the analysis of the specific electromagnetic signature to future work.  

%The possibles outcomes form a compact WD-MS binary are: cataclysmic variable \citep{Knigge2011}, novae \citep{Chomiuk2021}, Type Ia supernova progenitor (if the single degenerate scenario works). As mentioned before, in the context of this work we group all these outcome together. 

We adopt $m_1=0.6M_\odot$ and $m_2=1M_\odot$ for the primary WD and the secondary MS star, respectively. 
%For this channel, we set the following parameters. The masses $m_1=0.6M_\odot$ and $m_2=1M_\odot$, represent the mass of the primary WD and the secondary MS star, respectively.
In this case, the primary is the WD, even though it is the less massive component because the primary/secondary distinction is defined in the MS lifetime. $m_1$ represents the more massive MS star that evolved to be the WD sooner than its companion. From Kroupa IMF we find the fraction of the primary to be a WD. We limited the mass range to be between $3M_{\odot}-8M_{\odot}$. In this mass range, we assume the primary evolves to a WD in $10^8\rm yr$, resulting in a fraction of the primary $f_{m_1}=0.02$. Assuming a uniform distribution of the mass ratio $q\in \left(0.3,1\right)$, the fraction of the companion to be an MS star is  $f_{m_2}=0.56$.  The binary fraction for this mass scale is $f_{\rm binary}=0.5$. We emphasize here for simplicity we represent all WD and MS stars with a single mass of $0.6M_\odot$ for the MS and $1M_\odot$ for the MS star. The interaction radius, $d=1R_{\odot}$, represents the distance in which tidal interaction between the WD and the MS.  In \ref{fig:All_N_LogR} we present the number of systems of interest from WD-MS channel as a function of radius from the SMBH after Hubble time, $t=10^{10}\rm yr$ in purple dotted line. In Figure \ref{fig:All_N_logT} we present the total number of systems of interest in the purple dotted line as a function of time.

\subsection{\label{Sec:WD-WD}WD-WD and SN/ LISA sources}
 In the last channel, we present here the following parameters that represent the binary population of DWD. The masses are set to be $m_1=0.6M_\odot$ and $m_2=0.6M_\odot$. The interaction radius, $d=1R_{\odot}$, represents the distance in which tidal interaction between the binary WD. Similar to the previous channel we limit the MS mass range to be $3M_{\odot}-8M_{\odot}$, hence from Kroupa IMF we find  $f_{m_1}=0.02$, and under the assumption of uniform distribution of the mass ratio $q\in \left(0.3,1\right)$ the fraction of the companion to be a MS star is  $f_{m_2}=0.5$. Similar to the previous channel the binary fraction for this mass scale is $f_{\rm binary}=0.42$.

The possible outcomes from a compact DWD binary are: Type Ia supernova progenitor, as the two WD inspiral toward each other while omitting GWs and eventually merge and exceeding the Chandrasekhar limit which in turn leads to detonation of the merged product (if the double degenerate scenario works). Another way to achieve detonation in the binary systems is by a head-on collisions between the two WDs \citep[e.g.,][]{Katz2012,Kushnir2013}. 

As mentioned above inspiraling DWD are a source for millehertz GW signals. These signals, if close enough, i.e. in our galaxy, might be detectable by the future LISA mission  \citep{Marsh2011,Amaro-Seoane2012}. In Figure \ref{fig:All_N_LogR} we present the number of systems of interest from DWD channel as a function of radius from the SMBH after Hubble time, $t=10^{10}\rm yr$ in purple dotted line. In Figure \ref{fig:All_N_logT} we present the total number of systems of interest in the purple dotted line as a function of time.

\section{\label{sec:Discussion}Discussion}% and summary}

%\subsection{Spatial distribution}  %Spacial
%In what follows, we explain the shape of the spatial distribution presented in Figure \ref{fig:All_N_LogR}. 

Given the galactic center's extensive observational campaign \citep[e.g.,][]{Ghez2000,Gillessen2009,Stostd2015,Chen2023}, the spatial distribution of the interacting binaries can be used to constrain the star formation process in the inner $500$~pc. As depicted in Figure \ref{fig:All_N_LogR}, the probability of interaction at the binary's pericenter is a function of time and position in the galactic center.  We consider a single starburst model; thus, initially, the interacting binaries are localized closer to the SMBH ($\lsim 20$~pc). As time goes by (above $\sim$Gyr), the interacting binaries are localized at the outer parts ($lsim$few tens of parsec. Therefore, it is straightforward to understand the continuous or episodic star formation rate at this vicinity \citep[e.g.,][]{Lu2009,Chen2023}, from future observations of interacting binaries.

\begin{figure*}
\includegraphics[width=1\columnwidth] {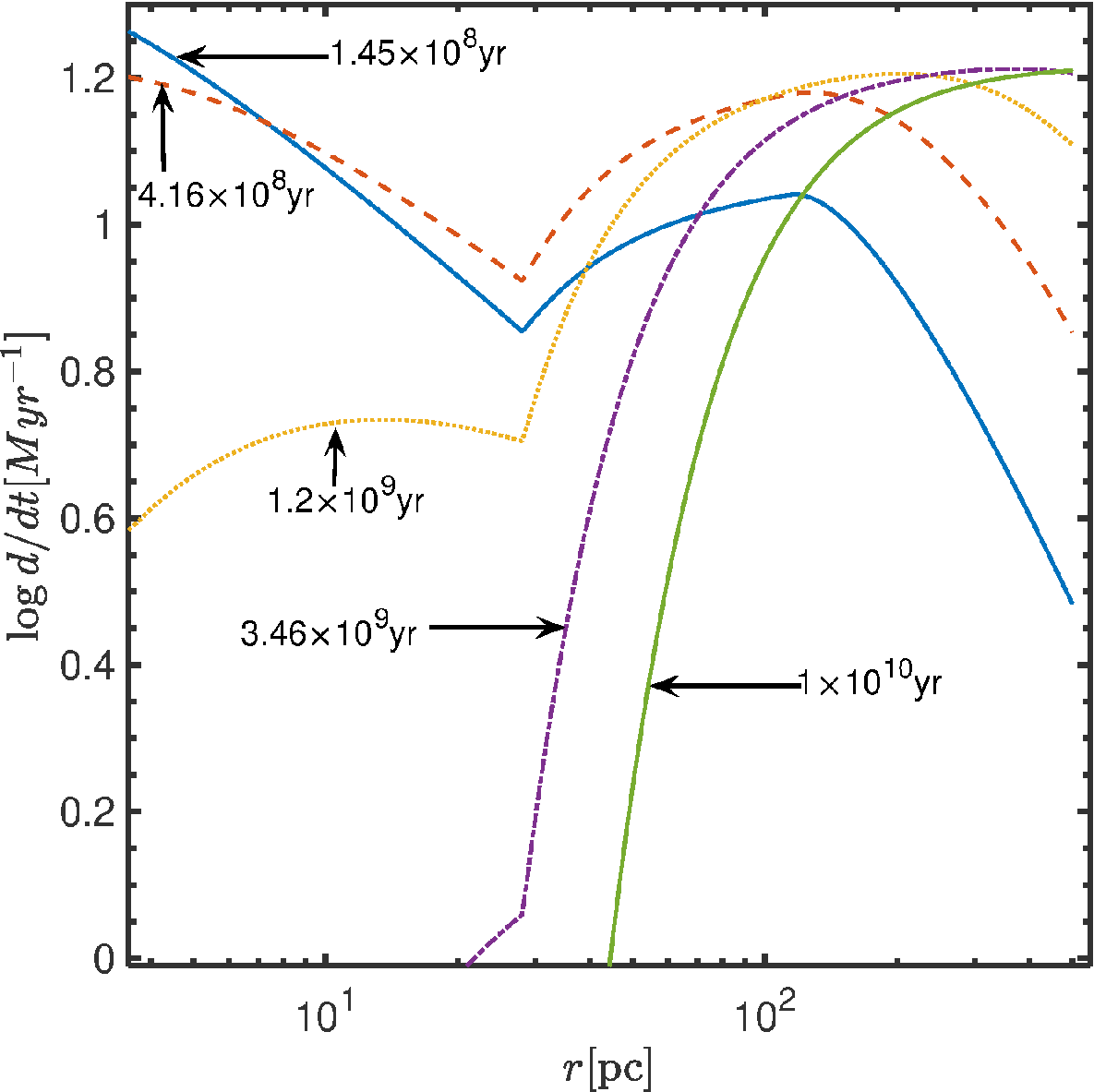} \qquad \includegraphics[width=1\columnwidth]{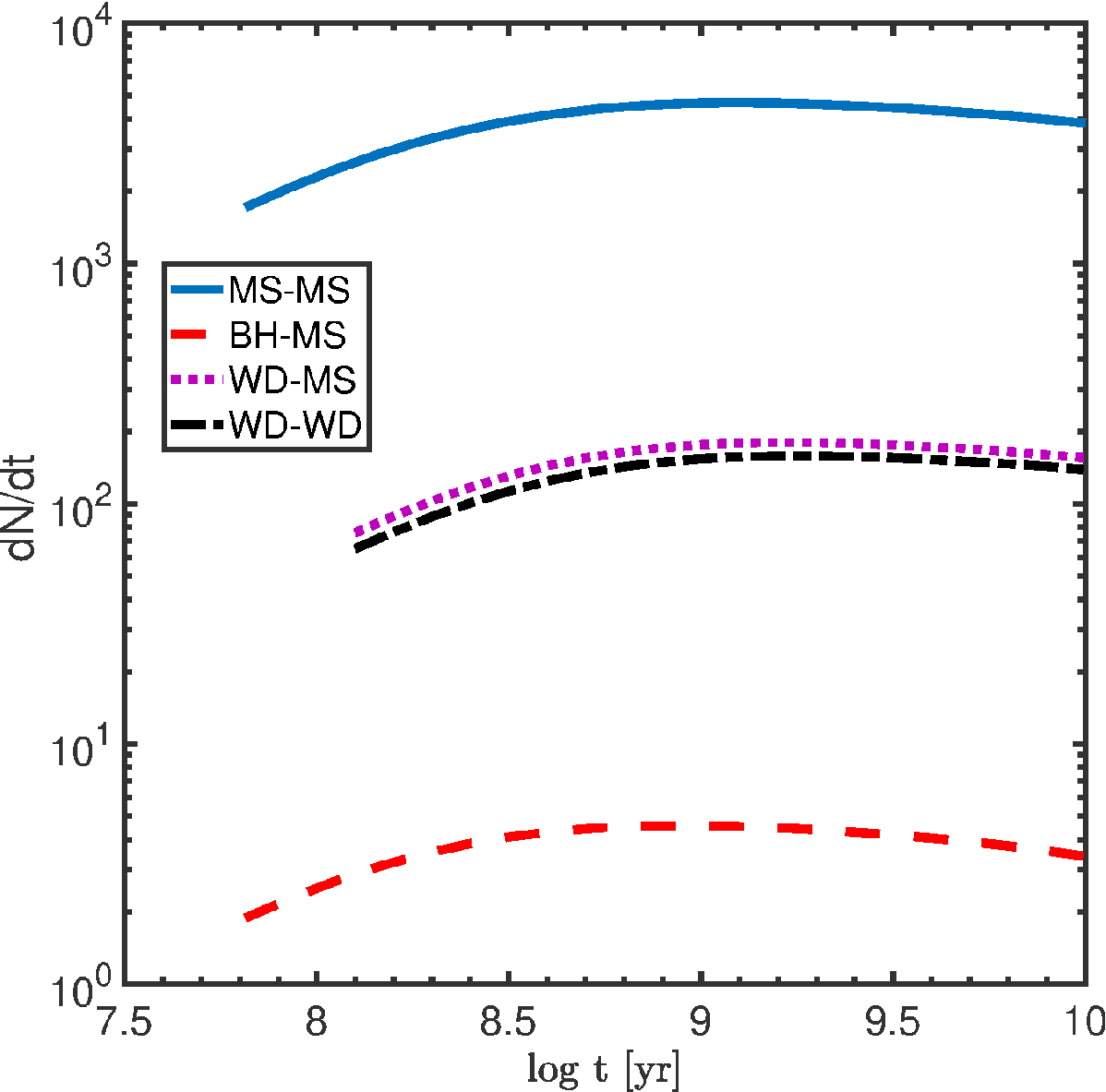}
\caption{\label{fig:DTD_BMS_different_times}Left panel, the differential contribution to the interacting binaries as a function of position for different times since formation. As the blue solid line is the closet to star formation it exhibits the highest contribution in the more denser regions with highest velocity dispersion. As time progress we see less and less contribution in the more dense regions and more contributions in the less dense region. The grin solid line presents the contribution of the last time step, there is no contribution from the inner region all "new" interactions happen closer to the edge of the field. Right panel, the delay time distribution for all channels considered. The color code is as Figure \ref{fig:All_N_LogR}. All channels exhibit the same peak of formation rate after $10^9 \rm yr$. }
\end{figure*}

The localization features and the peaks of the rate as a function of distance from the SMBH are straightforward to explain. Particularly, ignoring ionization, the probability of interaction is an increasing function of time. However, the ionization, which decreases the number of available binaries in this scenario, is extremely efficient in a high-density and high-velocity dispersion environment. As a result, we expect decreasing efficiency as a function of time, starting from the radius of influence and working its way to the edge of the Galactic field (heuristically referred to as $\sim 500$~pc). Simultaneously, the probability of interactions in the empty loss cone increases with higher velocity dispersion. Hence, we expect to see a high interaction rate at the radius of influence at early times, which decreases in space as $\sigma$ decreases. Once the velocity dispersion reaches its constant field value, the merger probability change comes only as a function of the local density via the critical SMA, $a_{\rm critical}$. These behaviors are depicted in Figure \ref{fig:DTD_BMS_different_times}.

%We now explain the peak all channels exhibit at $t=10^{10}\rm yr$ around $200\rm pc$. 
As shown in Figures \ref{fig:All_N_LogR} and  \ref{fig:DTD_BMS_different_times},  the number of systems and the derivative have a maximum at $\sim 200\rm pc$.  Particularly, the total number of systems interacting at the pericenter is the result of integrating the probability, (Eq.~(\ref{eq:Gamma})), twice, once over the binary's SMA and once over the binary position from the center. Thus, the boundaries of the interaction of the SMA are set by the lower cut-off, from impulse approximating considerations, and the upper cut-off, set by the ionization timescale (see Section \ref{sec:Physical Picture}).
%The total number of systems interacting at pericenter is the double integral on the probability, Eq.~(\ref{eq:Gamma}). The boundaries of the interaction of the SMA are set by the lower cut-off,from impulse approximating considerations, and the upper cut-off, set by the ionization time-scale. 
In the denser regions of the galactic center, the upper cut-off is smaller than the critical SMA, $a_{\rm critical}$. As the highest probability for interaction occurs for binaries with SMA equal to $a_{\rm critical}$. Thus, the total number of binaries increases as we approach the location in the galactic center that is adding the contribution of the full loss cone, i.e., binaries with $a>a_{\rm critical}$, this crossover occurs in $r \approx 200\rm pc$, see Figure \ref{fig:The-merger-probability} panel b.

The delay time distribution (DTD), $dN/dt$, is the rate of forming the systems of interest as a function of time since star formation. Specifically is the time derivative of Equation (\ref{eq:Gamma}) and the time derivative of Figure \ref{fig:All_N_logT}. In the previous subsection, we discussed the DTD as a function of position in the galactic center. In what follows, we present the DTD after integration on the distance from the SMBH so we are only considering the total number of interacting binaries as a function of time. We present the DTD for all channel studies in the letter in Figure \ref{fig:DTD_BMS_different_times}. The interaction probability peaks at $t=10^9\rm yr$ after star formation.

%We emphasize the fact that in this study we assume a single star burst at $t=0$, so if we consider continuous star formation we would have seen now the binary interactions of a star burst happened ~$1\rm Gyr$ ago.

\section{Summary\label{sec:Summary}}

Close binaries are often considered as a main source of exotic and transient phenomena Typically, studies focus on binaries in the field of galaxies \citep[e.g.,][and many others]{Iben1984,Kashi2010,Soker2012,ilkov2012,belczynski2002,andronov2006}, and even very close to an SMBH \citep[e.g.,][and others]{antonini2012,hoang2018,Stephan2019,hoang2020,wang2020,Rose2023}. However, the intermediate regime between the SMBH's radius of influence and the field. This unique regime has a dense environment such that soft binaries are collisional. Unlike the environment in the field, the stellar density is high. Furthermore, despite the high stellar density, these binaries remain soft over long timescales. Thus allowing them to interact with flyby stars, which can excite the binaries' eccentricities to high values \citep[][]{Michaely2016,michaely2019,michaely2020,Michaely2020b,Michaely2021,Raveh2022,michaely2022}. 

The mentioned unique environment yields a smaller binary separation, compared to the field, that maximizes the interaction at the pericenter due to flyby (see Figure \ref{fig:The-merger-probability} panel a). This separation depends on the location of the binary in the galactic center (as also depicted in Figure \ref{fig:The-merger-probability} panel b). Interestingly, inward to $\sim 100$~pc the binaries separation is limited by the ionization process. Thus, the binary ``wideness,'' depends on their location and varies from roughly $10-10,000$~au, for $\sim 4 - 500$~pc, respectively.

Here, we presented a novel dynamical channel that produces binary interaction from wide binaries in the galactic center. Wide binaries were thought not to play a role in binary stellar evolution due to their susceptibility to being ionized and because they were too wide to interact. However, we show that due to their larger cross-section, wide binaries tend to interact with local flyby stars. These interactions change the eccentricity and might gradually excite it to the point of interaction at the pericenter. 

As a proof of concept, we consider four interesting interacting binaries formed via this channel. Particularly we consider: 
%In order to present the importance of the channel we calculated the binary interaction rate for 4 different binary groups: 
MS-MS, BH-MS, WD-MS, and WD-WD. We choose these binaries due to the rich possible outcomes from their interactions. We show that the DTD peak at $t=10^9\rm yr$ and the location of the interaction gradually changes from the radius of influence, close to star formation, to the edge of the field as time progresses. We find that ~$10^5$ BMS in the galactic center interact due this process during the life-time of the MW galaxy. DWD and WD-MS interact ~$10^3$ of times and BH-MS ~$10^2$ of times.

\section*{Acknowledgments}
 EM and SN thank the Bhaumik Institute for Theoretical Physics and Howard and Astrid Preston for their generous support. We also acknowledge the partial support from NASA ATP 80NSSC20K0505 and from NSF-AST 2206428 
%%\end{Acknowledgments}

\bibliographystyle{aasjournal}

\begin{thebibliography}{}
\expandafter\ifx\csname natexlab\endcsname\relax\def\natexlab#1{#1}\fi
\providecommand{\url}[1]{\href{#1}{#1}}
\providecommand{\dodoi}[1]{doi:~\href{http://doi.org/#1}{\nolinkurl{#1}}}
\providecommand{\doeprint}[1]{\href{http://ascl.net/#1}{\nolinkurl{http://ascl.net/#1}}}
\providecommand{\doarXiv}[1]{\href{https://arxiv.org/abs/#1}{\nolinkurl{https://arxiv.org/abs/#1}}}

\bibitem[{{Aharon} \& {Perets}(2016)}]{Aharon2016}
{Aharon}, D., \& {Perets}, H.~B. 2016, \apjl, 830, L1,
  \dodoi{10.3847/2041-8205/830/1/L1}

\bibitem[{{Alexander}(2005)}]{Alexander2005}
{Alexander}, T. 2005, \physrep, 419, 65, \dodoi{10.1016/j.physrep.2005.08.002}

\bibitem[{{Amaro-Seoane} {et~al.}(2012){Amaro-Seoane}, {Aoudia}, {Babak},
  {Bin{\'e}truy}, {Berti}, {Boh{\'e}}, {Caprini}, {Colpi}, {Cornish},
  {Danzmann}, {Dufaux}, {Gair}, {Jennrich}, {Jetzer}, {Klein}, {Lang}, {Lobo},
  {Littenberg}, {McWilliams}, {Nelemans}, {Petiteau}, {Porter}, {Schutz},
  {Sesana}, {Stebbins}, {Sumner}, {Vallisneri}, {Vitale}, {Volonteri}, \&
  {Ward}}]{Amaro-Seoane2012}
{Amaro-Seoane}, P., {Aoudia}, S., {Babak}, S., {et~al.} 2012, Classical and
  Quantum Gravity, 29, 124016, \dodoi{10.1088/0264-9381/29/12/124016}

\bibitem[{{Andronov} {et~al.}(2006){Andronov}, {Pinsonneault}, \&
  {Terndrup}}]{andronov2006}
{Andronov}, N., {Pinsonneault}, M.~H., \& {Terndrup}, D.~M. 2006, \apj, 646,
  1160, \dodoi{10.1086/505127}

\bibitem[{{Antonini} {et~al.}(2010){Antonini}, {Faber}, {Gualandris}, \&
  {Merritt}}]{Antonini2010}
{Antonini}, F., {Faber}, J., {Gualandris}, A., \& {Merritt}, D. 2010, \apj,
  713, 90, \dodoi{10.1088/0004-637X/713/1/90}

\bibitem[{{Antonini} {et~al.}(2011){Antonini}, {Lombardi}, \&
  {Merritt}}]{Antonini2011}
{Antonini}, F., {Lombardi}, James~C., J., \& {Merritt}, D. 2011, \apj, 731,
  128, \dodoi{10.1088/0004-637X/731/2/128}

\bibitem[{{Antonini} \& {Perets}(2012)}]{antonini2012}
{Antonini}, F., \& {Perets}, H.~B. 2012, \apj, 757, 27,
  \dodoi{10.1088/0004-637X/757/1/27}

\bibitem[{{Athanassoula} {et~al.}(2017){Athanassoula}, {Rodionov}, \&
  {Prantzos}}]{Athanassoula2017}
{Athanassoula}, E., {Rodionov}, S.~A., \& {Prantzos}, N. 2017, \mnras, 467,
  L46, \dodoi{10.1093/mnrasl/slw255}

\bibitem[{{Bahcall} {et~al.}(1985){Bahcall}, {Hut}, \&
  {Tremaine}}]{Bahcall1985}
{Bahcall}, J.~N., {Hut}, P., \& {Tremaine}, S. 1985, \apj, 290, 15,
  \dodoi{10.1086/162953}

\bibitem[{{Ballero} {et~al.}(2007){Ballero}, {Matteucci}, {Origlia}, \&
  {Rich}}]{Ballero2007}
{Ballero}, S.~K., {Matteucci}, F., {Origlia}, L., \& {Rich}, R.~M. 2007, \aap,
  467, 123, \dodoi{10.1051/0004-6361:20066596}

\bibitem[{{Barbuy} {et~al.}(2018){Barbuy}, {Chiappini}, \&
  {Gerhard}}]{Barbuy2018}
{Barbuy}, B., {Chiappini}, C., \& {Gerhard}, O. 2018, \araa, 56, 223,
  \dodoi{10.1146/annurev-astro-081817-051826}

\bibitem[{{Belczynski} {et~al.}(2002){Belczynski}, {Kalogera}, \&
  {Bulik}}]{belczynski2002}
{Belczynski}, K., {Kalogera}, V., \& {Bulik}, T. 2002, \apj, 572, 407,
  \dodoi{10.1086/340304}

\bibitem[{{Carney} {et~al.}(2005){Carney}, {Latham}, \& {Laird}}]{carney2005}
{Carney}, B.~W., {Latham}, D.~W., \& {Laird}, J.~B. 2005, \aj, 129, 466,
  \dodoi{10.1086/426566}

\bibitem[{{Carney} {et~al.}(2001){Carney}, {Latham}, {Laird}, {Grant}, \&
  {Morse}}]{carney2001}
{Carney}, B.~W., {Latham}, D.~W., {Laird}, J.~B., {Grant}, C.~E., \& {Morse},
  J.~A. 2001, \aj, 122, 3419, \dodoi{10.1086/324233}

\bibitem[{{Chen} {et~al.}(2023){Chen}, {Do}, {Ghez}, {Hosek},
  {Feldmeier-Krause}, {Chu}, {Bentley}, {Lu}, \& {Morris}}]{Chen2023}
{Chen}, Z., {Do}, T., {Ghez}, A.~M., {et~al.} 2023, \apj, 944, 79,
  \dodoi{10.3847/1538-4357/aca8ad}

\bibitem[{{Chomiuk} {et~al.}(2021){Chomiuk}, {Metzger}, \&
  {Shen}}]{Chomiuk2021}
{Chomiuk}, L., {Metzger}, B.~D., \& {Shen}, K.~J. 2021, \araa, 59, 391,
  \dodoi{10.1146/annurev-astro-112420-114502}

\bibitem[{{Chu} {et~al.}(2018){Chu}, {Do}, {Hees}, {Ghez}, {Naoz}, {Witzel},
  {Sakai}, {Chappell}, {Gautam}, {Lu}, \& {Matthews}}]{Chu2018ApJ}
{Chu}, D.~S., {Do}, T., {Hees}, A., {et~al.} 2018, \apj, 854, 12,
  \dodoi{10.3847/1538-4357/aaa3eb}

\bibitem[{{Ciurlo} {et~al.}(2020){Ciurlo}, {Campbell}, {Morris}, {Do}, {Ghez},
  {Hees}, {Sitarski}, {Kosmo O'Neil}, {Chu}, {Martinez}, {Naoz}, \&
  {Stephan}}]{Ciurlo2020}
{Ciurlo}, A., {Campbell}, R.~D., {Morris}, M.~R., {et~al.} 2020, \nat, 577,
  337, \dodoi{10.1038/s41586-019-1883-y}

\bibitem[{{Di Stefano} {et~al.}(2023){Di Stefano}, {Kruckow}, {Gao},
  {Neunteufel}, \& {Kobayashi}}]{DiStefano2023}
{Di Stefano}, R., {Kruckow}, M.~U., {Gao}, Y., {Neunteufel}, P.~G., \&
  {Kobayashi}, C. 2023, \apj, 944, 87, \dodoi{10.3847/1538-4357/acae9b}

\bibitem[{{El-Badry} {et~al.}(2021){El-Badry}, {Rix}, \&
  {Heintz}}]{El-Badry2021}
{El-Badry}, K., {Rix}, H.-W., \& {Heintz}, T.~M. 2021, \mnras, 506, 2269,
  \dodoi{10.1093/mnras/stab323}

\bibitem[{{Ferrarese} \& {Ford}(2005)}]{Ferrarese2005}
{Ferrarese}, L., \& {Ford}, H. 2005, \ssr, 116, 523,
  \dodoi{10.1007/s11214-005-3947-6}

\bibitem[{{Fragione} {et~al.}(2019){Fragione}, {Grishin}, {Leigh}, {Perets}, \&
  {Perna}}]{fragione2019}
{Fragione}, G., {Grishin}, E., {Leigh}, N. W.~C., {Perets}, H.~B., \& {Perna},
  R. 2019, \mnras, 488, 47, \dodoi{10.1093/mnras/stz1651}

\bibitem[{{Gautam} {et~al.}(2019){Gautam}, {Do}, {Ghez}, {Morris}, {Martinez},
  {Hosek}, {Lu}, {Sakai}, {Witzel}, {Jia}, {Becklin}, \&
  {Matthews}}]{Gautam2019}
{Gautam}, A.~K., {Do}, T., {Ghez}, A.~M., {et~al.} 2019, \apj, 871, 103,
  \dodoi{10.3847/1538-4357/aaf103}

\bibitem[{{Gebhardt} {et~al.}(1996){Gebhardt}, {Richstone}, {Ajhar}, {Lauer},
  {Byun}, {Kormendy}, {Dressler}, {Faber}, {Grillmair}, \&
  {Tremaine}}]{Gebhardt1996}
{Gebhardt}, K., {Richstone}, D., {Ajhar}, E.~A., {et~al.} 1996, \aj, 112, 105,
  \dodoi{10.1086/117992}

\bibitem[{{Gebhardt} {et~al.}(2000{\natexlab{a}}){Gebhardt}, {Bender}, {Bower},
  {Dressler}, {Faber}, {Filippenko}, {Green}, {Grillmair}, {Ho}, {Kormendy},
  {Lauer}, {Magorrian}, {Pinkney}, {Richstone}, \& {Tremaine}}]{Gebhardt2000}
{Gebhardt}, K., {Bender}, R., {Bower}, G., {et~al.} 2000{\natexlab{a}}, \apjl,
  539, L13, \dodoi{10.1086/312840}

\bibitem[{{Gebhardt} {et~al.}(2000{\natexlab{b}}){Gebhardt}, {Kormendy}, {Ho},
  {Bender}, {Bower}, {Dressler}, {Faber}, {Filippenko}, {Green}, {Grillmair},
  {Lauer}, {Magorrian}, {Pinkney}, {Richstone}, \& {Tremaine}}]{Gebhardt2000b}
{Gebhardt}, K., {Kormendy}, J., {Ho}, L.~C., {et~al.} 2000{\natexlab{b}},
  \apjl, 543, L5, \dodoi{10.1086/318174}

\bibitem[{{Genzel} {et~al.}(2010){Genzel}, {Eisenhauer}, \&
  {Gillessen}}]{Genzel2010}
{Genzel}, R., {Eisenhauer}, F., \& {Gillessen}, S. 2010, Reviews of Modern
  Physics, 82, 3121, \dodoi{10.1103/RevModPhys.82.3121}

\bibitem[{Genzel {et~al.}(2003)Genzel, Schödel, Ott, Eisenhauer, Hofmann,
  Lehnert, Eckart, Alexander, Sternberg, Lenzen, Clénet, Lacombe, Rouan,
  Renzini, \& Tacconi-Garman}]{Genzel2003}
Genzel, R., Schödel, R., Ott, T., {et~al.} 2003, The Astrophysical Journal,
  594, 812, \dodoi{10.1086/377127}

\bibitem[{{Ghez} {et~al.}(1998){Ghez}, {Klein}, {Morris}, \&
  {Becklin}}]{Ghez1998}
{Ghez}, A.~M., {Klein}, B.~L., {Morris}, M., \& {Becklin}, E.~E. 1998, \apj,
  509, 678, \dodoi{10.1086/306528}

\bibitem[{{Ghez} {et~al.}(2000){Ghez}, {Morris}, {Becklin}, {Tanner}, \&
  {Kremenek}}]{Ghez2000}
{Ghez}, A.~M., {Morris}, M., {Becklin}, E.~E., {Tanner}, A., \& {Kremenek}, T.
  2000, \nat, 407, 349, \dodoi{10.1038/35030032}

\bibitem[{{Ghez} {et~al.}(2005){Ghez}, {Salim}, {Hornstein}, {Tanner}, {Lu},
  {Morris}, {Becklin}, \& {Duch{\^e}ne}}]{Ghez2005}
{Ghez}, A.~M., {Salim}, S., {Hornstein}, S.~D., {et~al.} 2005, \apj, 620, 744,
  \dodoi{10.1086/427175}

\bibitem[{{Ghez} {et~al.}(2003){Ghez}, {Duch{\^e}ne}, {Matthews}, {Hornstein},
  {Tanner}, {Larkin}, {Morris}, {Becklin}, {Salim}, {Kremenek}, {Thompson},
  {Soifer}, {Neugebauer}, \& {McLean}}]{Ghez2003}
{Ghez}, A.~M., {Duch{\^e}ne}, G., {Matthews}, K., {et~al.} 2003, \apjl, 586,
  L127, \dodoi{10.1086/374804}

\bibitem[{{Ghez} {et~al.}(2008){Ghez}, {Salim}, {Weinberg}, {Lu}, {Do}, {Dunn},
  {Matthews}, {Morris}, {Yelda}, {Becklin}, {Kremenek}, {Milosavljevic}, \&
  {Naiman}}]{2008Ghez}
{Ghez}, A.~M., {Salim}, S., {Weinberg}, N.~N., {et~al.} 2008, \apj, 689, 1044,
  \dodoi{10.1086/592738}

\bibitem[{{Gillessen} {et~al.}(2009){Gillessen}, {Eisenhauer}, {Trippe},
  {Alexander}, {Genzel}, {Martins}, \& {Ott}}]{Gillessen2009}
{Gillessen}, S., {Eisenhauer}, F., {Trippe}, S., {et~al.} 2009, \apj, 692,
  1075, \dodoi{10.1088/0004-637X/692/2/1075}

\bibitem[{{Gillessen} {et~al.}(2012){Gillessen}, {Genzel}, {Fritz}, {Quataert},
  {Alig}, {Burkert}, {Cuadra}, {Eisenhauer}, {Pfuhl}, {Dodds-Eden}, {Gammie},
  \& {Ott}}]{Gillessen2012}
{Gillessen}, S., {Genzel}, R., {Fritz}, T.~K., {et~al.} 2012, \nat, 481, 51,
  \dodoi{10.1038/nature10652}

\bibitem[{{Gillessen} {et~al.}(2019){Gillessen}, {Plewa}, {Widmann}, {von
  Fellenberg}, {Schartmann}, {Habibi}, {Jimenez Rosales}, {Baub{\"o}ck},
  {Dexter}, {Gao}, {Waisberg}, {Eisenhauer}, {Pfuhl}, {Ott}, {Burkert}, {de
  Zeeuw}, \& {Genzel}}]{Gillessen2019}
{Gillessen}, S., {Plewa}, P.~M., {Widmann}, F., {et~al.} 2019, \apj, 871, 126,
  \dodoi{10.3847/1538-4357/aaf4f8}

\bibitem[{{Grishin} \& {Perets}(2022)}]{Grishin2022}
{Grishin}, E., \& {Perets}, H.~B. 2022, \mnras, 512, 4993,
  \dodoi{10.1093/mnras/stac706}

\bibitem[{{Hawley}(2012)}]{Hawley2012}
{Hawley}, W.~P. 2012, PhD thesis, Arizona State University

\bibitem[{{Heggie}(1975)}]{Heggie1975}
{Heggie}, D.~C. 1975, \mnras, 173, 729, \dodoi{10.1093/mnras/173.3.729}

\bibitem[{{Heintz} {et~al.}(2022){Heintz}, {Hermes}, {El-Badry}, {Walsh}, {van
  Saders}, {Fields}, \& {Koester}}]{Heintz2022}
{Heintz}, T.~M., {Hermes}, J.~J., {El-Badry}, K., {et~al.} 2022, \apj, 934,
  148, \dodoi{10.3847/1538-4357/ac78d9}

\bibitem[{{Hoang} {et~al.}(2018){Hoang}, {Naoz}, {Kocsis}, {Rasio}, \&
  {Dosopoulou}}]{hoang2018}
{Hoang}, B.-M., {Naoz}, S., {Kocsis}, B., {Rasio}, F.~A., \& {Dosopoulou}, F.
  2018, \apj, 856, 140, \dodoi{10.3847/1538-4357/aaafce}

\bibitem[{{Hoang} {et~al.}(2020){Hoang}, {Naoz}, \& {Kremer}}]{hoang2020}
{Hoang}, B.-M., {Naoz}, S., \& {Kremer}, K. 2020, \apj, 903, 8,
  \dodoi{10.3847/1538-4357/abb66a}

\bibitem[{{Hoang} {et~al.}(2022){Hoang}, {Naoz}, \& {Sloneker}}]{Hoang2022}
{Hoang}, B.-M., {Naoz}, S., \& {Sloneker}, M. 2022, \apj, 934, 54,
  \dodoi{10.3847/1538-4357/ac7787}

\bibitem[{{Hopman}(2009)}]{Hopman2009}
{Hopman}, C. 2009, \apj, 700, 1933, \dodoi{10.1088/0004-637X/700/2/1933}

\bibitem[{{Hopman} \& {Alexander}(2005)}]{Hopman2005}
{Hopman}, C., \& {Alexander}, T. 2005, \apj, 629, 362, \dodoi{10.1086/431475}

\bibitem[{{Iben} \& {Tutukov}(1984)}]{Iben1984}
{Iben}, I., J., \& {Tutukov}, A.~V. 1984, \apjs, 54, 335,
  \dodoi{10.1086/190932}

\bibitem[{{Igoshev} {et~al.}(2020){Igoshev}, {Perets}, \&
  {Michaely}}]{Igoshev2020}
{Igoshev}, A.~P., {Perets}, H.~B., \& {Michaely}, E. 2020, \mnras, 494, 1448,
  \dodoi{10.1093/mnras/staa833}

\bibitem[{{Ilkov} \& {Soker}(2012)}]{ilkov2012}
{Ilkov}, M., \& {Soker}, N. 2012, \mnras, 419, 1695,
  \dodoi{10.1111/j.1365-2966.2011.19833.x}

\bibitem[{{Ivanova} {et~al.}(2013){Ivanova}, {Justham}, {Chen}, {De Marco},
  {Fryer}, {Gaburov}, {Ge}, {Glebbeek}, {Han}, {Li}, {Lu}, {Marsh},
  {Podsiadlowski}, {Potter}, {Soker}, {Taam}, {Tauris}, {van den Heuvel}, \&
  {Webbink}}]{ivanova2013}
{Ivanova}, N., {Justham}, S., {Chen}, X., {et~al.} 2013, \aapr, 21, 59,
  \dodoi{10.1007/s00159-013-0059-2}

\bibitem[{{Kaib} \& {Raymond}(2014)}]{Kaib2014}
{Kaib}, N.~A., \& {Raymond}, S.~N. 2014, \apj, 782, 60,
  \dodoi{10.1088/0004-637X/782/2/60}

\bibitem[{{Kashi} \& {Soker}(2010)}]{Kashi2010}
{Kashi}, A., \& {Soker}, N. 2010, ArXiv:1011.1222.
\newblock \doarXiv{1011.1222}

\bibitem[{{Katz} \& {Dong}(2012)}]{Katz2012}
{Katz}, B., \& {Dong}, S. 2012, ArXiv.
\newblock \doarXiv{1211.4584}

\bibitem[{{Knigge} {et~al.}(2011){Knigge}, {Baraffe}, \&
  {Patterson}}]{Knigge2011}
{Knigge}, C., {Baraffe}, I., \& {Patterson}, J. 2011, \apjs, 194, 28,
  \dodoi{10.1088/0067-0049/194/2/28}

\bibitem[{{Kormendy} \& {Ho}(2013)}]{Kormendy2013}
{Kormendy}, J., \& {Ho}, L.~C. 2013, \araa, 51, 511,
  \dodoi{10.1146/annurev-astro-082708-101811}

\bibitem[{{Kraus} \& {Hillenbrand}(2009)}]{Kraus2009}
{Kraus}, A.~L., \& {Hillenbrand}, L.~A. 2009, \apj, 704, 531,
  \dodoi{10.1088/0004-637X/704/1/531}

\bibitem[{{Kremer} {et~al.}(2022){Kremer}, {Lombardi}, {Lu}, {Piro}, \&
  {Rasio}}]{Kremer2022}
{Kremer}, K., {Lombardi}, J.~C., {Lu}, W., {Piro}, A.~L., \& {Rasio}, F.~A.
  2022, \apj, 933, 203, \dodoi{10.3847/1538-4357/ac714f}

\bibitem[{{Kremer} {et~al.}(2023){Kremer}, {Mockler}, {Piro}, \&
  {Lombardi}}]{Kremer2023}
{Kremer}, K., {Mockler}, B., {Piro}, A.~L., \& {Lombardi}, J.~C. 2023, \mnras,
  524, 6358, \dodoi{10.1093/mnras/stad2239}

\bibitem[{{Kroupa}(2001)}]{Kroupa2001}
{Kroupa}, P. 2001, \mnras, 322, 231, \dodoi{10.1046/j.1365-8711.2001.04022.x}

\bibitem[{{Kulkarni} {et~al.}(2007){Kulkarni}, {Ofek}, {Rau}, {Cenko},
  {Soderberg}, {Fox}, {Gal-Yam}, {Capak}, {Moon}, {Li}, {Filippenko}, {Egami},
  {Kartaltepe}, \& {Sanders}}]{Kulkarni2007}
{Kulkarni}, S.~R., {Ofek}, E.~O., {Rau}, A., {et~al.} 2007, \nat, 447, 458,
  \dodoi{10.1038/nature05822}

\bibitem[{{Kushnir} {et~al.}(2013){Kushnir}, {Katz}, {Dong}, {Livne}, \&
  {Fern{\'a}ndez}}]{Kushnir2013}
{Kushnir}, D., {Katz}, B., {Dong}, S., {Livne}, E., \& {Fern{\'a}ndez}, R.
  2013, \apjl, 778, L37, \dodoi{10.1088/2041-8205/778/2/L37}

\bibitem[{{Langer} {et~al.}(2017){Langer}, {Velusamy}, {Morris}, {Goldsmith},
  \& {Pineda}}]{Langer2017}
{Langer}, W.~D., {Velusamy}, T., {Morris}, M.~R., {Goldsmith}, P.~F., \&
  {Pineda}, J.~L. 2017, \aap, 599, A136, \dodoi{10.1051/0004-6361/201629497}

\bibitem[{{Leonard}(1989)}]{Leonard1989}
{Leonard}, P. J.~T. 1989, \aj, 98, 217, \dodoi{10.1086/115138}

\bibitem[{{Lightman} \& {Shapiro}(1977)}]{Lightman1977}
{Lightman}, A.~P., \& {Shapiro}, S.~L. 1977, \apj, 211, 244,
  \dodoi{10.1086/154925}

\bibitem[{{Linial} \& {Sari}(2022)}]{Linial2022}
{Linial}, I., \& {Sari}, R. 2022, \apj, 940, 101,
  \dodoi{10.3847/1538-4357/ac9bfd}

\bibitem[{{Lor{\'e}n-Aguilar} {et~al.}(2010){Lor{\'e}n-Aguilar}, {Isern}, \&
  {Garc{\'\i}a-Berro}}]{Loren-Aguilar2010}
{Lor{\'e}n-Aguilar}, P., {Isern}, J., \& {Garc{\'\i}a-Berro}, E. 2010, \mnras,
  406, 2749, \dodoi{10.1111/j.1365-2966.2010.16878.x}

\bibitem[{{Lu} {et~al.}(2013){Lu}, {Do}, {Ghez}, {Morris}, {Yelda}, \&
  {Matthews}}]{Lu2013}
{Lu}, J.~R., {Do}, T., {Ghez}, A.~M., {et~al.} 2013, \apj, 764, 155,
  \dodoi{10.1088/0004-637X/764/2/155}

\bibitem[{{Lu} {et~al.}(2009){Lu}, {Ghez}, {Hornstein}, {Morris}, {Becklin}, \&
  {Matthews}}]{Lu2009}
{Lu}, J.~R., {Ghez}, A.~M., {Hornstein}, S.~D., {et~al.} 2009, \apj, 690, 1463,
  \dodoi{10.1088/0004-637X/690/2/1463}

\bibitem[{{Magorrian} \& {Tremaine}(1999)}]{Magorrian1999}
{Magorrian}, J., \& {Tremaine}, S. 1999, \mnras, 309, 447,
  \dodoi{10.1046/j.1365-8711.1999.02853.x}

\bibitem[{{Marsh}(2011)}]{Marsh2011}
{Marsh}, T.~R. 2011, Classical and Quantum Gravity, 28, 094019,
  \dodoi{10.1088/0264-9381/28/9/094019}

\bibitem[{{Martins} \& {Plez}(2006)}]{Martins2006}
{Martins}, F., \& {Plez}, B. 2006, \aap, 457, 637,
  \dodoi{10.1051/0004-6361:20065753}

\bibitem[{{Merritt}(2013)}]{Merritt2013}
{Merritt}, D. 2013, Classical and Quantum Gravity, 30, 244005,
  \dodoi{10.1088/0264-9381/30/24/244005}

\bibitem[{{Merritt} \& {Poon}(2004)}]{Merritt2004}
{Merritt}, D., \& {Poon}, M.~Y. 2004, \apj, 606, 788, \dodoi{10.1086/382497}

\bibitem[{{Michaely}(2020)}]{Michaely2020b}
{Michaely}, E. 2020, \mnras, \dodoi{10.1093/mnras/staa3623}

\bibitem[{{Michaely}(2021)}]{michaely2021a}
---. 2021, \mnras, 500, 5543, \dodoi{10.1093/mnras/staa3623}

\bibitem[{{Michaely} \& {Naoz}(2022)}]{michaely2022}
{Michaely}, E., \& {Naoz}, S. 2022, \apj, 936, 184,
  \dodoi{10.3847/1538-4357/ac8a92}

\bibitem[{{Michaely} \& {Perets}(2016)}]{Michaely2016}
{Michaely}, E., \& {Perets}, H.~B. 2016, \mnras, 458, 4188,
  \dodoi{10.1093/mnras/stw368}

\bibitem[{{Michaely} \& {Perets}(2019{\natexlab{a}})}]{michaely2019}
---. 2019{\natexlab{a}}, \apjl, 887, L36, \dodoi{10.3847/2041-8213/ab5b9b}

\bibitem[{{Michaely} \& {Perets}(2019{\natexlab{b}})}]{michaely2019a}
---. 2019{\natexlab{b}}, \mnras, 484, 4711, \dodoi{10.1093/mnras/stz352}

\bibitem[{Michaely \& Perets(2020)}]{michaely2020}
Michaely, E., \& Perets, H.~B. 2020, Monthly Notices of the Royal Astronomical
  Society, \dodoi{10.1093/mnras/staa2720}

\bibitem[{{Michaely} \& {Shara}(2021)}]{Michaely2021}
{Michaely}, E., \& {Shara}, M.~M. 2021, \mnras, 502, 4540,
  \dodoi{10.1093/mnras/stab339}

\bibitem[{{Mills}(2017)}]{Mills2017}
{Mills}, E.~A.~C. 2017, arXiv e-prints, arXiv:1705.05332,
  \dodoi{10.48550/arXiv.1705.05332}

\bibitem[{{Moe} \& {Di Stefano}(2016)}]{Moe2016}
{Moe}, M., \& {Di Stefano}, R. 2016, ArXiv e-prints.
\newblock \doarXiv{1606.05347}

\bibitem[{{Muno} {et~al.}(2006){Muno}, {Law}, {Clark}, {Dougherty}, {de Grijs},
  {Portegies Zwart}, \& {Yusef-Zadeh}}]{Muno2006}
{Muno}, M.~P., {Law}, C., {Clark}, J.~S., {et~al.} 2006, \apj, 650, 203,
  \dodoi{10.1086/507175}

\bibitem[{{Muno} {et~al.}(2009){Muno}, {Bauer}, {Baganoff}, {Bandyopadhyay},
  {Bower}, {Brandt}, {Broos}, {Cotera}, {Eikenberry}, {Garmire}, {Hyman},
  {Kassim}, {Lang}, {Lazio}, {Law}, {Mauerhan}, {Morris}, {Nagata},
  {Nishiyama}, {Park}, {Ram{\`\i}rez}, {Stolovy}, {Wijnands}, {Wang}, {Wang},
  \& {Yusef-Zadeh}}]{Muno2009}
{Muno}, M.~P., {Bauer}, F.~E., {Baganoff}, F.~K., {et~al.} 2009, \apjs, 181,
  110, \dodoi{10.1088/0067-0049/181/1/110}

\bibitem[{{Naoz}(2016)}]{Naoz2016}
{Naoz}, S. 2016, \araa, 54, 441, \dodoi{10.1146/annurev-astro-081915-023315}

\bibitem[{{Naoz} {et~al.}(2018){Naoz}, {Ghez}, {Hees}, {Do}, {Witzel}, \&
  {Lu}}]{Naoz2018}
{Naoz}, S., {Ghez}, A.~M., {Hees}, A., {et~al.} 2018, \apjl, 853, L24,
  \dodoi{10.3847/2041-8213/aaa6bf}

\bibitem[{{Naoz} {et~al.}(2022){Naoz}, {Rose}, {Michaely}, {Melchor},
  {Ramirez-Ruiz}, {Mockler}, \& {Schnittman}}]{Naoz2022}
{Naoz}, S., {Rose}, S.~C., {Michaely}, E., {et~al.} 2022, \apjl, 927, L18,
  \dodoi{10.3847/2041-8213/ac574b}

\bibitem[{{Neumayer} {et~al.}(2020){Neumayer}, {Seth}, \&
  {B{\"o}ker}}]{Neumayer2020}
{Neumayer}, N., {Seth}, A., \& {B{\"o}ker}, T. 2020, \aapr, 28, 4,
  \dodoi{10.1007/s00159-020-00125-0}

\bibitem[{{Ott} {et~al.}(1999){Ott}, {Eckart}, \& {Genzel}}]{Ott1999}
{Ott}, T., {Eckart}, A., \& {Genzel}, R. 1999, \apj, 523, 248,
  \dodoi{10.1086/307712}

\bibitem[{{Pastorello} {et~al.}(2019){Pastorello}, {Mason}, {Taubenberger},
  {Fraser}, {Cortini}, {Tomasella}, {Botticella}, {Elias-Rosa}, {Kotak},
  {Smartt}, {Benetti}, {Cappellaro}, {Turatto}, {Tartaglia}, {Djorgovski},
  {Drake}, {Berton}, {Briganti}, {Brimacombe}, {Bufano}, {Cai}, {Chen},
  {Christensen}, {Ciabattari}, {Congiu}, {Dimai}, {Inserra}, {Kankare},
  {Magill}, {Maguire}, {Martinelli}, {Morales-Garoffolo}, {Ochner}, {Pignata},
  {Reguitti}, {Sollerman}, {Spiro}, {Terreran}, \& {Wright}}]{Pastorello2019}
{Pastorello}, A., {Mason}, E., {Taubenberger}, S., {et~al.} 2019, \aap, 630,
  A75, \dodoi{10.1051/0004-6361/201935999}

\bibitem[{{Pejcha} {et~al.}(2016{\natexlab{a}}){Pejcha}, {Metzger}, \&
  {Tomida}}]{Pejcha2016}
{Pejcha}, O., {Metzger}, B.~D., \& {Tomida}, K. 2016{\natexlab{a}}, \mnras,
  455, 4351, \dodoi{10.1093/mnras/stv2592}

\bibitem[{{Pejcha} {et~al.}(2016{\natexlab{b}}){Pejcha}, {Metzger}, \&
  {Tomida}}]{Pejcha2016b}
---. 2016{\natexlab{b}}, \mnras, 461, 2527, \dodoi{10.1093/mnras/stw1481}

\bibitem[{{Perets} \& {Fabrycky}(2009)}]{perets2009}
{Perets}, H.~B., \& {Fabrycky}, D.~C. 2009, \apj, 697, 1048,
  \dodoi{10.1088/0004-637X/697/2/1048}

\bibitem[{{Perets} {et~al.}(2016){Perets}, {Li}, {Lombardi}, \&
  {Milcarek}}]{Perets2016}
{Perets}, H.~B., {Li}, Z., {Lombardi}, Jr., J.~C., \& {Milcarek}, Jr., S.~R.
  2016, \apj, 823, 113, \dodoi{10.3847/0004-637X/823/2/113}

\bibitem[{{Pfuhl} {et~al.}(2014){Pfuhl}, {Alexander}, {Gillessen}, {Martins},
  {Genzel}, {Eisenhauer}, {Fritz}, \& {Ott}}]{Pfuhl2014}
{Pfuhl}, O., {Alexander}, T., {Gillessen}, S., {et~al.} 2014, \apj, 782, 101,
  \dodoi{10.1088/0004-637X/782/2/101}

\bibitem[{{Prodan} {et~al.}(2015){Prodan}, {Antonini}, \&
  {Perets}}]{Prodan2015}
{Prodan}, S., {Antonini}, F., \& {Perets}, H.~B. 2015, \apj, 799, 118,
  \dodoi{10.1088/0004-637X/799/2/118}

\bibitem[{{Rafelski} {et~al.}(2007){Rafelski}, {Ghez}, {Hornstein}, {Lu}, \&
  {Morris}}]{Rafelski2007}
{Rafelski}, M., {Ghez}, A.~M., {Hornstein}, S.~D., {Lu}, J.~R., \& {Morris}, M.
  2007, \apj, 659, 1241, \dodoi{10.1086/512062}

\bibitem[{{Raghavan} {et~al.}(2010){Raghavan}, {McAlister}, {Henry}, {Latham},
  {Marcy}, {Mason}, {Gies}, {White}, \& {ten Brummelaar}}]{raghavan2010}
{Raghavan}, D., {McAlister}, H.~A., {Henry}, T.~J., {et~al.} 2010, \apjs, 190,
  1, \dodoi{10.1088/0067-0049/190/1/1}

\bibitem[{{Raskin} {et~al.}(2010){Raskin}, {Scannapieco}, {Rockefeller},
  {Fryer}, {Diehl}, \& {Timmes}}]{Raskin2010}
{Raskin}, C., {Scannapieco}, E., {Rockefeller}, G., {et~al.} 2010, \apj, 724,
  111, \dodoi{10.1088/0004-637X/724/1/111}

\bibitem[{{Raskin} {et~al.}(2009){Raskin}, {Timmes}, {Scannapieco}, {Diehl}, \&
  {Fryer}}]{Raskin2009}
{Raskin}, C., {Timmes}, F.~X., {Scannapieco}, E., {Diehl}, S., \& {Fryer}, C.
  2009, \mnras, 399, L156, \dodoi{10.1111/j.1745-3933.2009.00743.x}

\bibitem[{{Raveh} {et~al.}(2022){Raveh}, {Michaely}, \& {Perets}}]{Raveh2022}
{Raveh}, Y., {Michaely}, E., \& {Perets}, H.~B. 2022, \mnras, 514, 4246,
  \dodoi{10.1093/mnras/stac1605}

\bibitem[{{Rom} {et~al.}(2023){Rom}, {Linial}, \& {Sari}}]{Rom2023}
{Rom}, B., {Linial}, I., \& {Sari}, R. 2023, \apj, 951, 14,
  \dodoi{10.3847/1538-4357/acd54f}

\bibitem[{{R{\"o}pke} \& {De Marco}(2023)}]{Ropke2023}
{R{\"o}pke}, F.~K., \& {De Marco}, O. 2023, Living Reviews in Computational
  Astrophysics, 9, 2, \dodoi{10.1007/s41115-023-00017-x}

\bibitem[{{Rose} {et~al.}(2020){Rose}, {Naoz}, {Gautam}, {Ghez}, {Do}, {Chu},
  \& {Becklin}}]{Rose2020}
{Rose}, S.~C., {Naoz}, S., {Gautam}, A.~K., {et~al.} 2020, \apj, 904, 113,
  \dodoi{10.3847/1538-4357/abc557}

\bibitem[{{Rose} {et~al.}(2023){Rose}, {Naoz}, {Sari}, \& {Linial}}]{Rose2023}
{Rose}, S.~C., {Naoz}, S., {Sari}, R., \& {Linial}, I. 2023, \apj, 955, 30,
  \dodoi{10.3847/1538-4357/acee75}

\bibitem[{{Rosswog} {et~al.}(2009){Rosswog}, {Kasen}, {Guillochon}, \&
  {Ramirez-Ruiz}}]{Rosswog2009}
{Rosswog}, S., {Kasen}, D., {Guillochon}, J., \& {Ramirez-Ruiz}, E. 2009,
  \apjl, 705, L128, \dodoi{10.1088/0004-637X/705/2/L128}

\bibitem[{{Rozner} \& {Perets}(2023)}]{Rozner2023}
{Rozner}, M., \& {Perets}, H.~B. 2023, arXiv e-prints, arXiv:2304.02029,
  \dodoi{10.48550/arXiv.2304.02029}

\bibitem[{{Sana} {et~al.}(2014){Sana}, {Le Bouquin}, {Lacour}, {Berger},
  {Duvert}, {Gauchet}, {Norris}, {Olofsson}, {Pickel}, {Zins}, {Absil}, {de
  Koter}, {Kratter}, {Schnurr}, \& {Zinnecker}}]{Sana2014}
{Sana}, H., {Le Bouquin}, J.-B., {Lacour}, S., {et~al.} 2014, \apjs, 215, 15,
  \dodoi{10.1088/0067-0049/215/1/15}

\bibitem[{{Sari} \& {Fragione}(2019)}]{Sari2019}
{Sari}, R., \& {Fragione}, G. 2019, \apj, 885, 24,
  \dodoi{10.3847/1538-4357/ab43df}

\bibitem[{{Shara}(1999)}]{shara1999}
{Shara}, M.~M. 1999, \physrep, 311, 363, \dodoi{10.1016/S0370-1573(98)00115-X}

\bibitem[{{Shara} \& {Regev}(1986)}]{Shara1986}
{Shara}, M.~M., \& {Regev}, O. 1986, \apj, 306, 543, \dodoi{10.1086/164364}

\bibitem[{{Shara} \& {Shaviv}(1977)}]{Shara1977}
{Shara}, M.~M., \& {Shaviv}, G. 1977, \mnras, 179, 705,
  \dodoi{10.1093/mnras/179.4.705}

\bibitem[{{Shara} \& {Shaviv}(1978)}]{Shara1978}
---. 1978, \mnras, 183, 687, \dodoi{10.1093/mnras/183.4.687}

\bibitem[{{Shariat} {et~al.}(2023){Shariat}, {Naoz}, {Hansen}, {Angelo},
  {Michaely}, \& {Stephan}}]{Shariat2023}
{Shariat}, C., {Naoz}, S., {Hansen}, B. M.~S., {et~al.} 2023, arXiv e-prints,
  arXiv:2306.13130, \dodoi{10.48550/arXiv.2306.13130}

\bibitem[{{Soker}(2019)}]{soker2019}
{Soker}, N. 2019, \nar, 87, 101535, \dodoi{10.1016/j.newar.2020.101535}

\bibitem[{{Soker} \& {Kashi}(2012)}]{Soker2012}
{Soker}, N., \& {Kashi}, A. 2012, \apj, 746, 100,
  \dodoi{10.1088/0004-637X/746/1/100}

\bibitem[{{Sormani} {et~al.}(2020){Sormani}, {Tress}, {Glover}, {Klessen},
  {Battersby}, {Clark}, {Hatchfield}, \& {Smith}}]{Sormani2020}
{Sormani}, M.~C., {Tress}, R.~G., {Glover}, S. C.~O., {et~al.} 2020, \mnras,
  497, 5024, \dodoi{10.1093/mnras/staa1999}

\bibitem[{{Stephan} {et~al.}(2016){Stephan}, {Naoz}, {Ghez}, {Witzel},
  {Sitarski}, {Do}, \& {Kocsis}}]{Stephan2016}
{Stephan}, A.~P., {Naoz}, S., {Ghez}, A.~M., {et~al.} 2016, \mnras, 460, 3494,
  \dodoi{10.1093/mnras/stw1220}

\bibitem[{{Stephan} {et~al.}(2019){Stephan}, {Naoz}, {Ghez}, {Morris},
  {Ciurlo}, {Do}, {Breivik}, {Coughlin}, \& {Rodriguez}}]{Stephan2019}
---. 2019, \apj, 878, 58, \dodoi{10.3847/1538-4357/ab1e4d}

\bibitem[{{St{\o}stad} {et~al.}(2015){St{\o}stad}, {Do}, {Murray}, {Lu},
  {Yelda}, \& {Ghez}}]{Stostd2015}
{St{\o}stad}, M., {Do}, T., {Murray}, N., {et~al.} 2015, \apj, 808, 106,
  \dodoi{10.1088/0004-637X/808/2/106}

\bibitem[{{Tokovinin}(1997)}]{Tokovinin1997}
{Tokovinin}, A.~A. 1997, \aaps, 124, 75, \dodoi{10.1051/aas:1997181}

\bibitem[{{Wang} {et~al.}(2020){Wang}, {Stephan}, {Naoz}, {Hoang}, \&
  {Breivik}}]{wang2020}
{Wang}, H., {Stephan}, A.~P., {Naoz}, S., {Hoang}, B.-M., \& {Breivik}, K.
  2020, arXiv e-prints, arXiv:2010.15841.
\newblock \doarXiv{2010.15841}

\bibitem[{{Webbink}(1984)}]{Webbink1984}
{Webbink}, R.~F. 1984, \apj, 277, 355, \dodoi{10.1086/161701}

\bibitem[{{Webbink} {et~al.}(1987){Webbink}, {Livio}, {Truran}, \&
  {Orio}}]{Webbink1987}
{Webbink}, R.~F., {Livio}, M., {Truran}, J.~W., \& {Orio}, M. 1987, \apj, 314,
  653, \dodoi{10.1086/165095}

\bibitem[{{Witzel} {et~al.}(2014){Witzel}, {Ghez}, {Morris}, {Sitarski},
  {Boehle}, {Naoz}, {Campbell}, {Becklin}, {Canalizo}, {Chappell}, {Do}, {Lu},
  {Matthews}, {Meyer}, {Stockton}, {Wizinowich}, \& {Yelda}}]{Witzel2014}
{Witzel}, G., {Ghez}, A.~M., {Morris}, M.~R., {et~al.} 2014, \apjl, 796, L8,
  \dodoi{10.1088/2041-8205/796/1/L8}

\bibitem[{{Witzel} {et~al.}(2017){Witzel}, {Sitarski}, {Ghez}, {Morris},
  {Hees}, {Do}, {Lu}, {Naoz}, {Boehle}, {Martinez}, {Chappell}, {Sch{\"o}del},
  {Meyer}, {Yelda}, {Becklin}, \& {Matthews}}]{Witzel2017}
{Witzel}, G., {Sitarski}, B.~N., {Ghez}, A.~M., {et~al.} 2017, \apj, 847, 80,
  \dodoi{10.3847/1538-4357/aa80ea}

\bibitem[{{Xin} {et~al.}(2023){Xin}, {Haiman}, {Perna}, {Wang}, \&
  {Ryu}}]{Xin2023}
{Xin}, C., {Haiman}, Z., {Perna}, R., {Wang}, Y., \& {Ryu}, T. 2023, arXiv
  e-prints, arXiv:2303.12846, \dodoi{10.48550/arXiv.2303.12846}

\end{thebibliography}



\end{document}